\documentclass[a4paper, amsfonts, amssymb, amsmath, reprint, showkeys, nofootinbib, twoside,superscriptaddress]{revtex4-2}
\usepackage[english]{babel}
\usepackage[utf8]{inputenc}
\usepackage[colorinlistoftodos, color=green!40, prependcaption]{todonotes}
\usepackage[left=15mm,right=15mm,top=20mm,columnsep=15pt]{geometry} 
\usepackage[pdftex, pdftitle={Article}, pdfauthor={Author}]{hyperref} % For hyperlinks in the PDF
\usepackage{balance}
\usepackage{soul}
\usepackage{xfrac}
\usepackage{color}
\usepackage{tabularx}

\begin{document}
\title{Automatic Aberration Correction for Transcranial Functional and Super-Resolution Ultrasound Imaging in Rodents and Nonhuman Primates}

%Automatic Closed-Loop Aberration Correction for Transcranial Functional and Super-Resolution Ultrasound Imaging of Brain Hemodynamics in Rodents and Nonhuman Primates

\author{Paul Xing}
\affiliation{Department of Engineering Physics, Polytechnique Montréal, Montreal, Canada}
\author{Antoine Malescot}
\affiliation{Department of Physiology and Pharmacology, Université de Montréal, Montreal, Canada}
\affiliation{Department of Stomatology, Université de Montréal, Université de Montréal, Montreal, Canada}
%\author{Adan Ulises Dominguez-Vargas}
%\affiliation{Département de Neurosciences, Faculté de Médecine, Université de Montréal, Montreal, Canada}
\author{Eric Martineau}
\affiliation{Department of Physiology and Pharmacology, Université de Montréal, Montreal, Canada}
\affiliation{Department of Stomatology, Université de Montréal, Université de Montréal, Montreal, Canada}
\author{Stephan Quessy}
\affiliation{Department of Neuroscience, Université de Montréal, Montreal, Canada}
\author{Ravi L. Rungta}
\affiliation{Department of Stomatology, Université de Montréal, Université de Montréal, Montreal, Canada}
\affiliation{Department of Neuroscience, Université de Montréal, Montreal, Canada}
\affiliation{Centre interdisciplinaire de recherche sur le cerveau et l’apprentissage (CIRCA), Université de Montréal, Montreal, Canada}
\author{Numa Dancause} 
\affiliation{Département de Neurosciences, Faculté de Médecine, Université de Montréal, Montreal, Canada}
\affiliation{Centre interdisciplinaire de recherche sur le cerveau et l’apprentissage (CIRCA), Université de Montréal, Montreal, Canada}
\author{Jean Provost}
\email[Correspondence email address: ]{paul.xing@polymtl.ca, jean.provost@polymtl.ca}% Your name
\affiliation{Department of Engineering Physics, Polytechnique Montréal, Montreal, Canada}
\affiliation{Montreal Heart Institute, Montreal, Canada}

\date{\today}

\begin{abstract}
%Skull-induced aberration is a major drawback of transcranial ultrasound localization microscopy (ULM). Microbubble detection is reduced, and false detections or localization errors can lead to resolution degradation and imaging artifacts, such as disconnected or duplicated vessels. The advent of deep learning led to the release of multiple open-source libraries for automatic gradient computation used in neural networks. In this work, we present a differentiable beamforming framework for automatic aberration correction in transcranial Doppler and ULM imaging of the brain. We proposed a spatially distributed delay-based parameterization of the aberration that can be optimized in a closed-loop manner using angular coherence as an objective function. We demonstrated transcranial ULM using differentiable beamforming \textit{in vivo} improved the resolution in the mouse brain and in the nonhuman primate (NHP) brain. We also showed that differentiable beamforming improved the sensitivity of transcranial functional ultrasound imaging (fUS) of the mouse brain, as well as improved hemodynamic quantification of ULM in mice. As a proof of concept, we extended differentiable beamforming to 3D transcranial ULM imaging of the NHP brain and showed that we were able to correct for aberration and remove artifacts, such as duplication of vessels. Those results demonstrate that differentiable beamforming could be a promising step toward non-invasive transcranial super-resolution and functional imaging of the brain.

Skull-induced aberrations remain a major drawback of transcranial ultrasound localization microscopy (ULM), degrading sensitivity and spatial accuracy through microbubble mislocalization, false detections, and imaging artifacts, such as disconnected or duplicated vessels. Here, we present a differentiable beamforming framework for automatic aberration correction in transcranial Doppler and ULM. Our approach uses spatially distributed delay-based parameterization of the aberration that is optimized in a closed-loop manner using angular coherence as an objective function. We demonstrate robust improvements of transcranial ULM,  \textit{in vivo}, with enhanced resolution of both mouse and nonhuman primate (NHP) brains. We also extended differentiable beamforming to functional measurements, with improvements in the sensitivity of transcranial functional ultrasound (fUS) and  ULM based hemodynamic quantification. Extending this approach to 3D transcranial ULM imaging in NHPs, we show efficient correction of skull induced aberrations and removal of artifacts, such as vessel duplications. By providing a fully automated and generalizable solution for aberration correction, this work lowers a major technical barrier to transcranial ultrasound imaging, enabling broader adoption of non-invasive, super-resolution and functional neuroimaging across laboratories and across species.

\end{abstract}


\maketitle

\section{Introduction}

Brain function and integrity are highly dependent on the adequate blood supply provided by a vast vascular network composed of arteries, arterioles, capillaries, venules, and veins. In addition to strokes, increasing evidence is showing that cerebral blood flow alterations are associated with cognitive impairment and neurodegenerative diseases, including Alzheimer's disease and dementia \cite{iadecola2017neurovascular, anderle2025vascular}. However, transcranial imaging allowing for the non-invasive \textit{in vivo} visualization of neurovascular coupling and hemodynamics in the brain microvasculature remains challenging. Imaging techniques currently used in clinical settings, such as computed tomography angiography (CTA) and magnetic resonance imaging (MRI), can provide anatomical maps of the brain vasculature with a resolution limited in the millimeter range \cite{uecker2010real, lin2009basic}. Due to their limited sensitivity to blood flow, conventional clinical ultrasound imaging techniques such as transcranial Doppler ultrasonography (TCD), are only reliable in the large basal arteries of the brain \cite{aaslid1982noninvasive}.

Ultrasound localization microscopy (ULM) \cite{christensen2014vivo, errico2015ultrafast} is a recently developed imaging technique that can overcome the diffraction limit of ultrasound. ULM is based on microbubbles injected into the bloodstream that can be individually localized and then tracked over time as they flow through the vasculature. By detecting these microbubble centers with subwavelength precision, ULM can achieve a resolution 10 times higher than conventional ultrasound techniques and imaging vessels as small as 10 $\mu$m. ULM has been utilized to image the brain microvasculature in different animal models, from rodents \cite{errico2015ultrafast} to sheep \cite{Coudert2024, bureau2025ultrasound} and nonhuman primates (NHPs) \cite{xing20253d}, as well as in humans \cite{demene2021transcranial}. ULM was successfully used to detect vascular changes occurring in Alzheimer's disease \cite{lowerison2024super, lin2024super}, aging \cite{lowerison2022aging}, strokes \cite{chavignon20223d, regensburger2024ultrasound}, aneurysms \cite{demene2021transcranial} or arteriovenous malformations \cite{schwarz2025ultrasound}. Recent developments in dynamic ULM (dULM) \cite{bourquin2022vivo, ghigo2024dynamic} and functional ULM (fULM) \cite{renaudin2022functional} allowed the measurement of brain-wide temporal hemodynamic changes in the microvasculature, such as pulsatility or neurovascular activity following somatosensorial or visual stimulation. Hence, ULM could play a major role in the understanding of cerebrovascular functions \textit{in vivo} and in depth \cite{lee2026assessing, thalgott2026non}.

However, skull-induced aberration remains a major drawback for transcranial ULM imaging of the brain. Microbubble signals are decreased in the presence of the skull, leading to a reduced microbubble detection sensitivity and a decrease in contrast and resolution for ULM images. In addition, aberration can affect localization accuracy \cite{soulioti2019super, mccall2021characterization} and cause major imaging artifacts such as vessel duplication \cite{robin2023vivo, xing2024phase, xing2025inverse, Coudert2024, bureau2025ultrasound}, as well as bias in velocity estimation \cite{xing2024phase, xing2025inverse}. Hence, performing aberration correction in transcranial ULM is critical to achieve reliable super-resolution vascular maps and hemodynamic quantification. Previous works on aberration correction techniques include using single point-like reflectors \cite{o1988phase, flax1988phase} or optimizing image quality metrics such as spatial coherence \cite{mallart1994adaptive} and speckle brightness \cite{nock1989phase} to perform adaptive focusing. Several methods have also been proposed to correct aberrations of the skull \cite{ivancevich2008real,lindsey20143, schoen2019heterogeneous}, but are limited mainly to the larger brain vessels. The use of microbubbles as guide stars \cite{demene2021transcranial, robin2023vivo} allowed transcranial ULM imaging in humans, but suffers from the need to detect isolated microbubbles and required imaging through the temporal bone, where the skull is thinner. The concept of ultrasound matrix imaging (UMI) for aberration correction \cite{lambert2022bultrasound, bureau2023three} has also recently been extended to ULM \cite{bureau2025ultrasound}. However, UMI requires an extensive number of transmit angles to perform aberration correction, which can reduce the effective frame rate of ULM. Due to the numerous challenges to image through the skull, multiple works on transcranial ULM do not perform aberration correction \cite{renaudin2022functional, chavignon20223d, demeulenaere2022vivo}.

%Our group previously proposed a model-based approach to retrieve the aberration function from the sparsity of the ultrasound signal that can be used even at high microbubble concentrations \cite{xing2025inverse}, but still relies on detection of microbubbles. 

The advent of deep learning led to the development and release of multiple open source frameworks and libraries for automatic gradient computation, such as PyTorch, TensorFlow, or Google JAX. Since then, convolutional neural networks have been widely deployed in speech recognition and computer-vision tasks. Deep learning strategies have also been successfully applied in different fields of ultrasound imaging \cite{van2019deep} as well as for the specific case of ULM \cite{rauby2024deep} and transcranial aberration correction \cite{xing2024phase, zhou2024transcranial}. 

%Complex-valued convolutional neural networks have been applied in rodents to correct for aberration in transcranial ULM \cite{xing2024phase}. 

However, the capability of neural networks to generalize beyond the training dataset can be limited, hindering the general implementation of aberration correction strategies in different settings. Recently, the concept of differentiable beamforming has been proposed to improve ultrasound focusing \cite{simson2023differentiable, heriard2025path}. Taking advantage of these automatic differentiation frameworks developed for deep learning, differentiable beamforming can correct for aberration by computing speed of sound maps from synthetic aperture data. Such approaches offer the advantage of a physics-informed framework to solve optimization problems via the same gradient-based strategies used in deep learning without the need of prior training or the use of an extensive dataset.

Inspired by these previous works, we present in this study a differentiable beamforming framework for transcranial imaging of the brain. We propose a spatially distributed delay-based parameterization to correct the aberration induced by the skull. We exploit the angular coherence of the ultrasound signal to retrieve the aberration function using gradient ascent. We successfully performed transcranial Doppler and ULM imaging in adult rodent and nonhuman primate (NHP) brains with improved contrast and resolution. We also performed transcranial functional ultrasound (fUS) imaging \cite{mace2011functional} of the mouse brain and showed that Doppler sensitivity increased after aberration correction. Finally, we demonstrated that differentiable beamforming can be extended to 3D transcranial ULM of the NHP brain, achieving aberration correction with a limited set of plane waves. This work establishes that differentiable beamforming overcomes a major barrier to transcranial ultrasound imaging by providing a generalizable solution to aberration correction and represents a promising step toward high-resolution, non-invasive imaging of brain structure and function.

\begin{figure*}[ht!]
    \centering
    \includegraphics[width=1\linewidth]{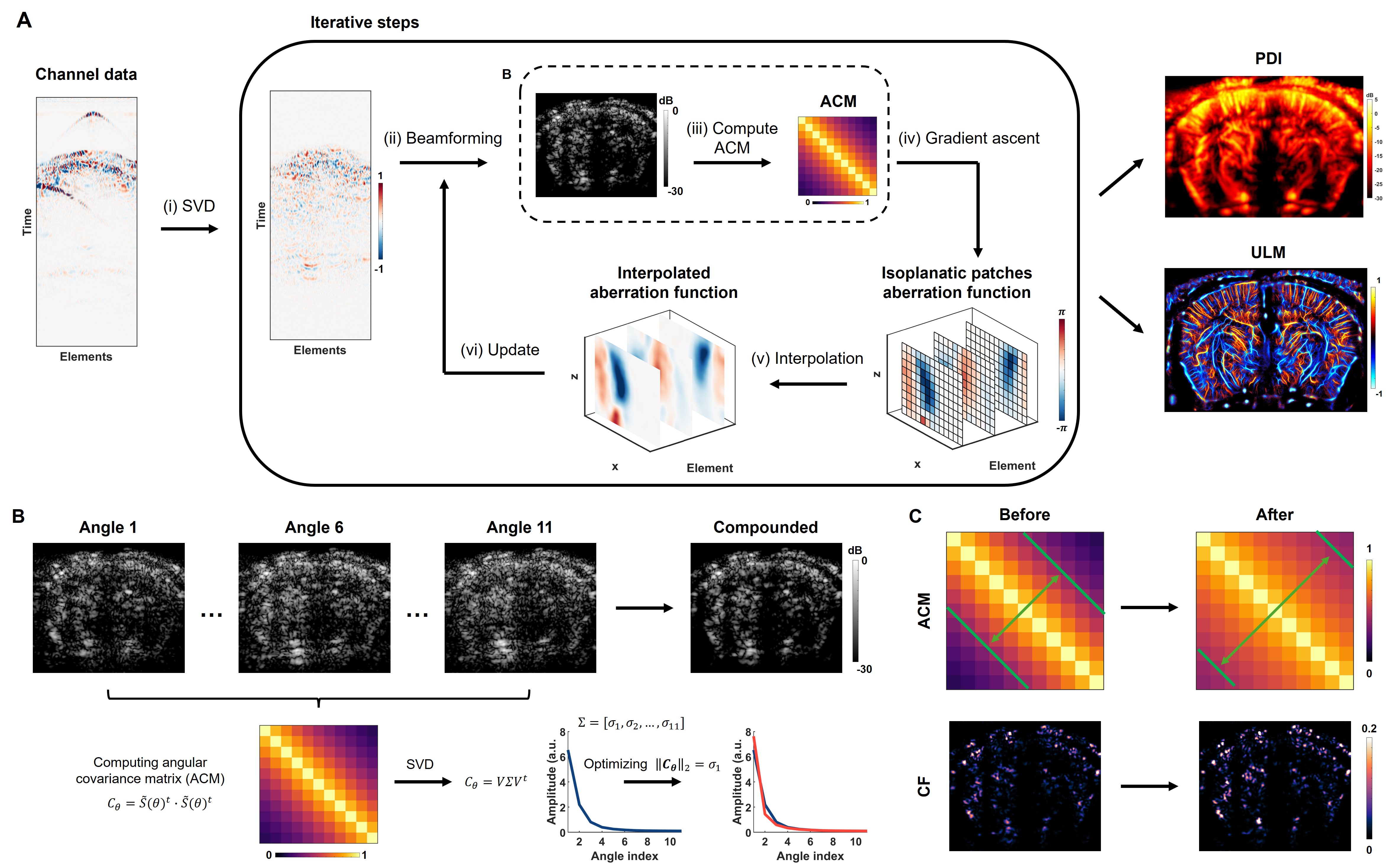}
    \caption[Differentiable beamforming for transcranial imaging of the brain]{Differentiable beamforming for transcranial imaging of the brain. (\textbf{A}) Main steps of the differentiable beamforming pipeline. (i) Channel data are SVD filtered prior to differentiable computation for tissue cancellation and to remove signal from the skull. (ii) Channel data are beamformed using a spatially-distributed phase delays function as model for the skull-induced aberration. (iii) The first singular value $\sigma_1$ of $\mathbf{C}_\theta$ is used as an objective function. (iv) The objective function is then differentiated with respect to the aberration function. The phase delays of each isoplanatic patch are then updated through gradient ascent. (v) The aberration function is interpolated to cover the entire beamforming grid. (vi) During the next iteration, the channel data are beamformed again using the updated version of the aberration function. These steps can be repeated iteratively until convergence or after a fixed number of iterations. (\textbf{B}) $\mathbf{C}_\theta$ is computed and its SVD decomposition is performed to retrieve the first singular value $\sigma_1$. Optimizing the value $\sigma_1$ is equivalent to optimizing the coherence between each plane wave image.
    (\textbf{C}) Examples of $\mathbf{C}_\theta$ and spatial CF map obtained before and after aberration correction. Magnitude of $\mathbf{C}_\theta$ and CF map are increased with differentiable beamforming, showing higher angular and spatial coherence after aberration correction.}
    \label{fig:methods}
\end{figure*}

\begin{figure*}
    \centering
    \includegraphics[width=1\linewidth]{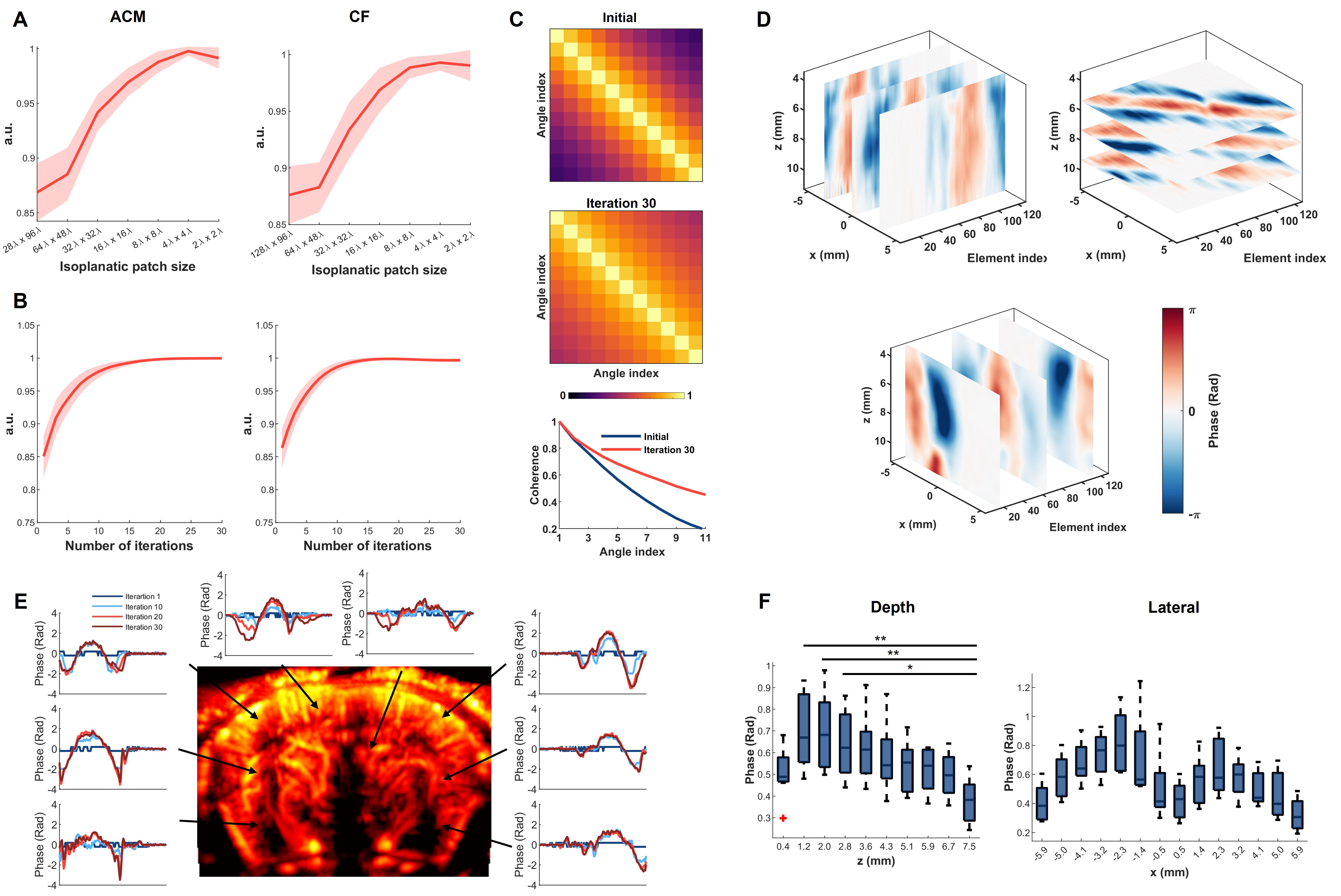}
    \caption[Differentiable beamforming convergence analysis]{Differentiable beamforming convergence analysis. (\textbf{A}) Normalized $\mathbf{C}_\theta$ ($n=7$ mice, $F=83.9$, $P<$ 0.0001, repeated-measures one-way ANOVA) and CF values ($n=7$ mice, $F=49.8$, $P<$ 0.0001, repeated-measures one-way ANOVA) for different numbers of isoplanatic patches. (\textbf{B}) Evolution of  $\mathbf{C}_\theta$ and CF across the number of iterations. (\textbf{C}) Example of  $\mathbf{C}_\theta$ matrix after aberration correction with off-diagonal profile. (\textbf{D}) Examples of aberration function obtained after 30 iterations. (\textbf{E}) Aberration profile for the different isoplanatic patches obtained after 1, 10, 20, and 30 iterations. (\textbf{F}) Depth and lateral variation of the aberration function ($n=7$ mice, $F=3.65$, $P<$ 0.01, two-way ANOVA with post-hoc \textit{t}-test and Bonferroni correction)}
    \label{fig:metrics}
\end{figure*}

\begin{figure*}
    \centering
    \includegraphics[width=1\linewidth]{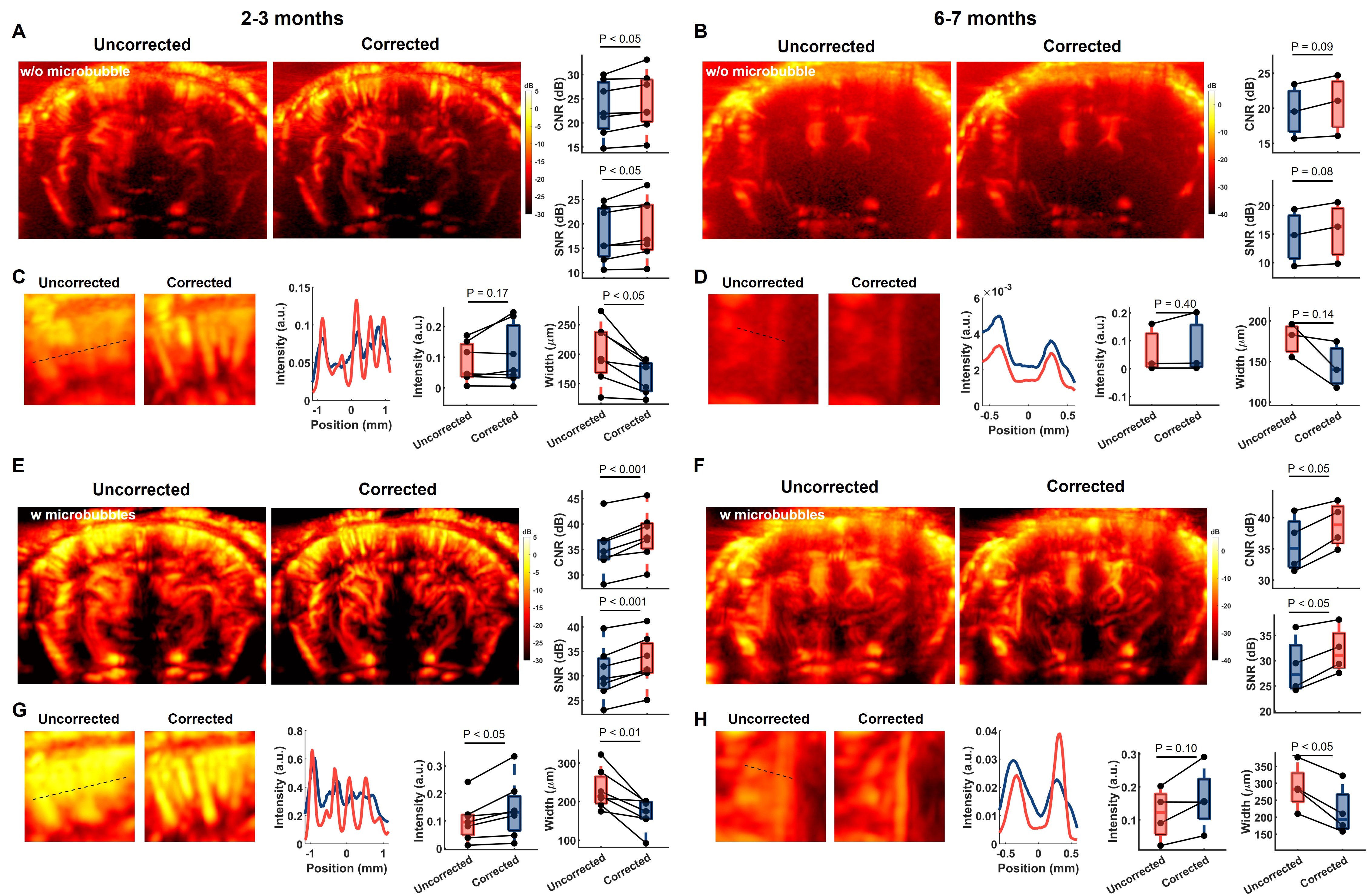}
    \caption[Transcranial power Doppler ultrasound imaging of the mouse brain with aberration correction]{Transcranial power Doppler ultrasound imaging of the mouse brain with aberration correction. (\textbf{A}) Aberration correction of PDI in young mice (2-3 months old) without use of microbubbles showed improvement in CNR ($n=7$, $P  < $ 0.05, two-tailed paired Student's \textit{t}-test) and SNR ($n=7$, $P  < $ 0.05, two-tailed paired Student's \textit{t}-test). (\textbf{B}) In older mice (6-7 months old), CNR ($n=3$, $P  = $ 0.09, two-tailed paired Student's \textit{t}-test) and SNR ($n=3$, $P  = $ 0.08, two-tailed paired Student's \textit{t}-test) tended to be improved after correction. (\textbf{C},\textbf{D}) Examples of vessel profile views before and after correction. Vessel intensity tended to increase after correction in young ($n=7$, $P  = $ 0.17, two-tailed paired Student's \textit{t}-test) and old mice ($n=3$, $P  = $ 0.40, two-tailed paired Student's \textit{t}-test). Vessel width improved in young ($n=7$ mice, $P  < $ 0.05, two-tailed paired Student's \textit{t}-test) and tended to improve in old mice ($n=3$ mice, $P  = $ 0.14, two-tailed paired Student's \textit{t}-test) after correction.
    (\textbf{E}) Aberration correction of CEUS PDI in young mice showed improvement in CNR ($n=7$, $P  < $ 0.001, two-tailed paired Student's \textit{t}-test) and SNR ($n=7$, $P  < $ 0.001, two-tailed paired Student's \textit{t}-test). (\textbf{F}) PDI in older mice (6-7 months old) showed improvement in CNR ($n=4$, $P  < $ 0.05, two-tailed paired Student's \textit{t}-test) and SNR ($n=4$, $P  < $ 0.05, two-tailed paired Student's \textit{t}-test). (\textbf{G},\textbf{H}) Example of vessel profile views before and after correction. Vessel intensity increased after correction in young mice ($n=7$, $P  < $ 0.05, two-tailed paired Student's \textit{t}-test) and tended to increase in old mice ($n=4$, $P  = $ 0.10, two-tailed paired Student's \textit{t}-test). Vessel width improved in young ($n=7$ mice, $P  < $ 0.01, two-tailed paired Student's \textit{t}-test) and in old mice ($n=3$ mice, $P  < $ 0.05, two-tailed paired Student's \textit{t}-test) after correction.}
    \label{fig:pdi}
\end{figure*}

\begin{figure*}
    \centering
    \includegraphics[width=1\linewidth]{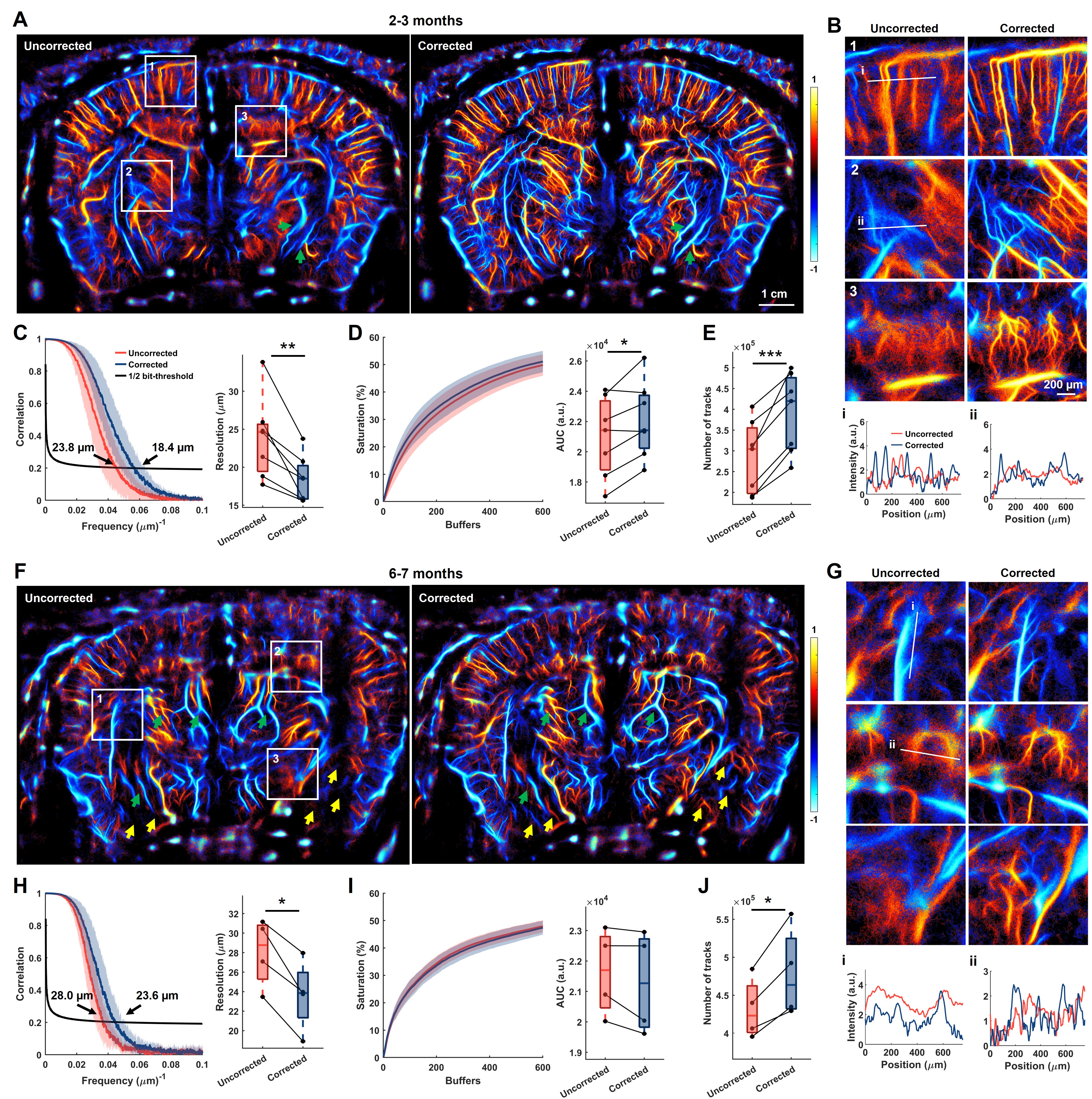}
    \caption[Transcranial ULM of the brain with aberration correction]{Transcranial ULM of the brain with aberration correction (\textbf{A}-\textbf{E}) in young and (\textbf{F}-\textbf{J}) old mice. (\textbf{A}, \textbf{F}) ULM maps obtained before and after correction. Green arrows indicate presence of disjoint vessels that are properly connected after aberration correction while yellow arrows indicate vessel recovered in shadowed areas. (\textbf{B}, \textbf{G}) Regions of interest (ROIs) with profile view examples. (\textbf{C}, \textbf{H}) Mean FRC curve with standard deviation. Resolution improved from 23.8 $\pm$ $5.4$ $\mu$m to 18.4 $\pm$ 3.0 $\mu$m after correction in young mice ($n=7$, $P  < $ 0.01, two-tailed paired Student's \textit{t}-test) and improved from 28.0 $\pm$ 3.5 $\mu$m to 23.6 $\pm$ 3.7 $\mu$m in old mice ($n=4$, $P  < $ 0.05, two-tailed paired Student's \textit{t}-test). 
    (\textbf{D}, \textbf{I}) Mean saturation curve with standard deviation. Area under the curve (AUC) increased from 2.10 $\pm$  $0.26\times10^4$ to  2.21 $\pm$ $0.25\times10^4$ after correction in young mice ($n=7$, $P  < $ 0.05, two-tailed paired Student's \textit{t}-test) but remained unchanged in old mice ($n=4$, $P  = $ 0.15, two-tailed paired Student's \textit{t}-test) . (\textbf{E}, \textbf{J}) The number of tracks increased from 2.84 $\pm$ $0.88\times10^5$ to  3.90 $\pm$  $0.96\times10^5$ after correction in young mice ($n=7$, $P  < $ 0.0001, two-tailed paired Student's \textit{t}-test) and increased from 4.32 $\pm$ $0.40\times10^5$ to  4.78 $\pm$  $0.60\times10^5$ in old mice ($n=4$, $P  < $ 0.05, two-tailed paired Student's \textit{t}-test). } 
    \label{fig:ULM_mice}
\end{figure*}
%    (\textbf{D}) FRC curves ($n=3$), (i) FRC resolution ($n=3$, $P  < $ \hl{XXX}, two-tailed paired Student's \textit{t}-test), and (j) saturation curves before and after correction in old mice.

\begin{figure*}
    \centering
    \includegraphics[width=0.9\linewidth]{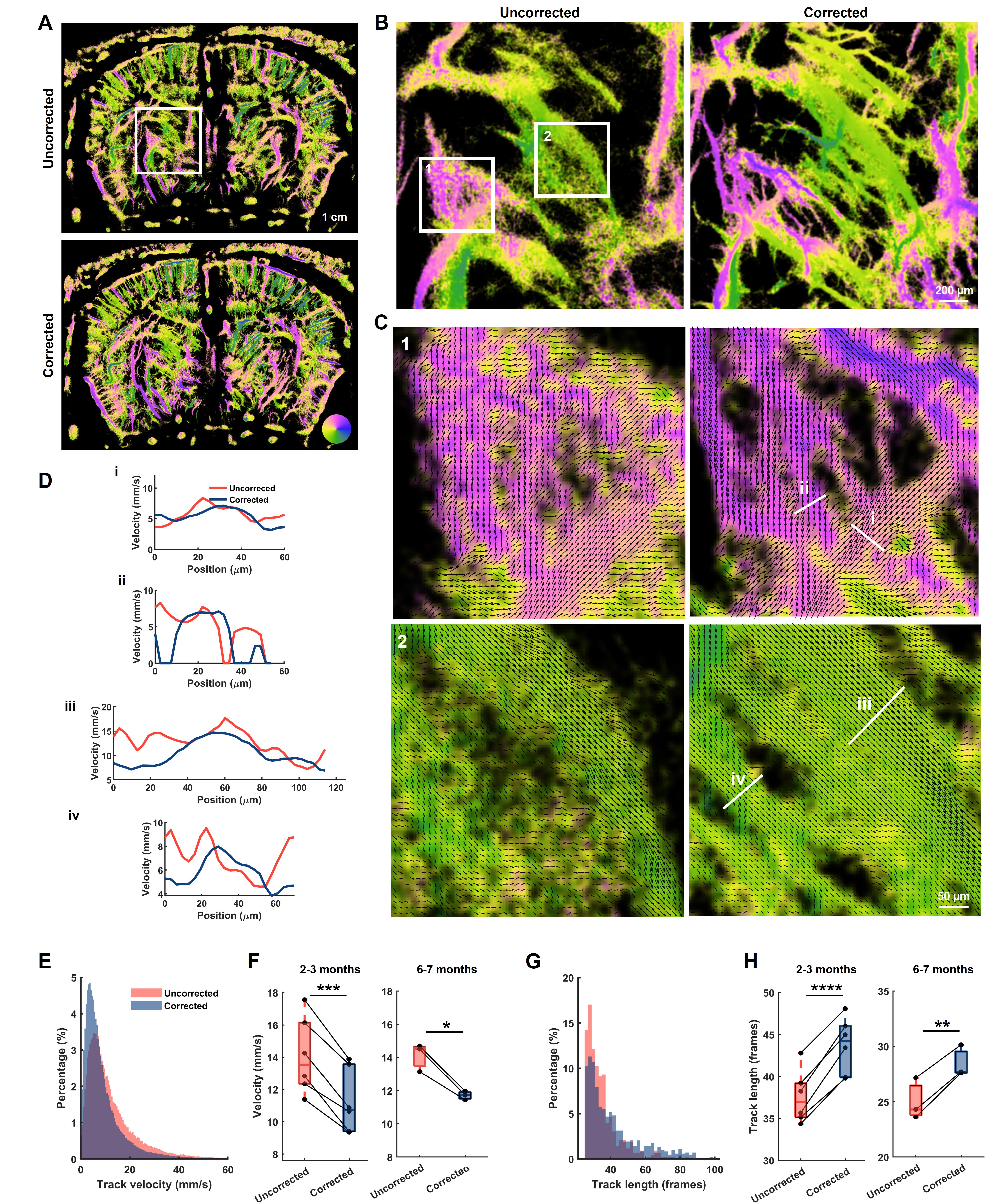}
    \caption[Aberration correction improves hemodynamic characterization using ULM]{Aberration correction improves hemodynamic characterization using ULM. (\textbf{A}) Flow direction map.  (\textbf{B}) ROIs extracted from (\textbf{A}). (\textbf{C}) Close-ups with flow vector fields.  (\textbf{D}) Velocity profile across vessels section. (\textbf{E}) track velocity distribution ($P<$ 0.0001 Kolmogorov–Smirnov test). (\textbf{F}) Mean track velocity before and after correction decreased from 14.1 $\pm$ 2.4 mm/s to from 11.3 $\pm$ 2.0 mm/s in young mice ($n=7$, $P  < $ 0.001, two-tailed paired Student's \textit{t}-test) and decreased from 14.1 $\pm$ 0.8 mm/s to 11.7 $\pm$ 0.3 mm/s in old mice ($n=4$, $P  < $ 0.05, two-tailed paired Student's \textit{t}-test). (\textbf{G}) track length distribution ($P<$ 0.0001 Kolmogorov–Smirnov test). (\textbf{H}) Mean track length before and after correction increased from 37.6 $\pm$ 2.4 to from 43.7 $\pm$ 3.4 in young mice ($n=7$, $P  < $ 0.0001, two-tailed paired Student's \textit{t}-test) and increased from 25.0 $\pm$ 1.9 frames to 28.5 $\pm$ 1.5 frames in old mice ($n=4$, $P  < $ 0.01, two-tailed paired Student's \textit{t}-test). }
    \label{fig:ulm_hemodynamics}
\end{figure*}

%(b) Velocity profile with quantitative assessment across vessels section ($P<$ \hl{XX}, ANOVA with post-hoc \textit{t}-test and Bonferroni correction). (d) track velocity distribution ($P<$ \hl{XX} Kolmogorov–Smirnov test) and (e) tracks counts for different velocities. (f) Track length distribution ($P<$ \hl{XX} Kolmogorov–Smirnov test)  and (g) track counts for different lengths.

\begin{figure*}
    \centering
    \includegraphics[width=1\linewidth]{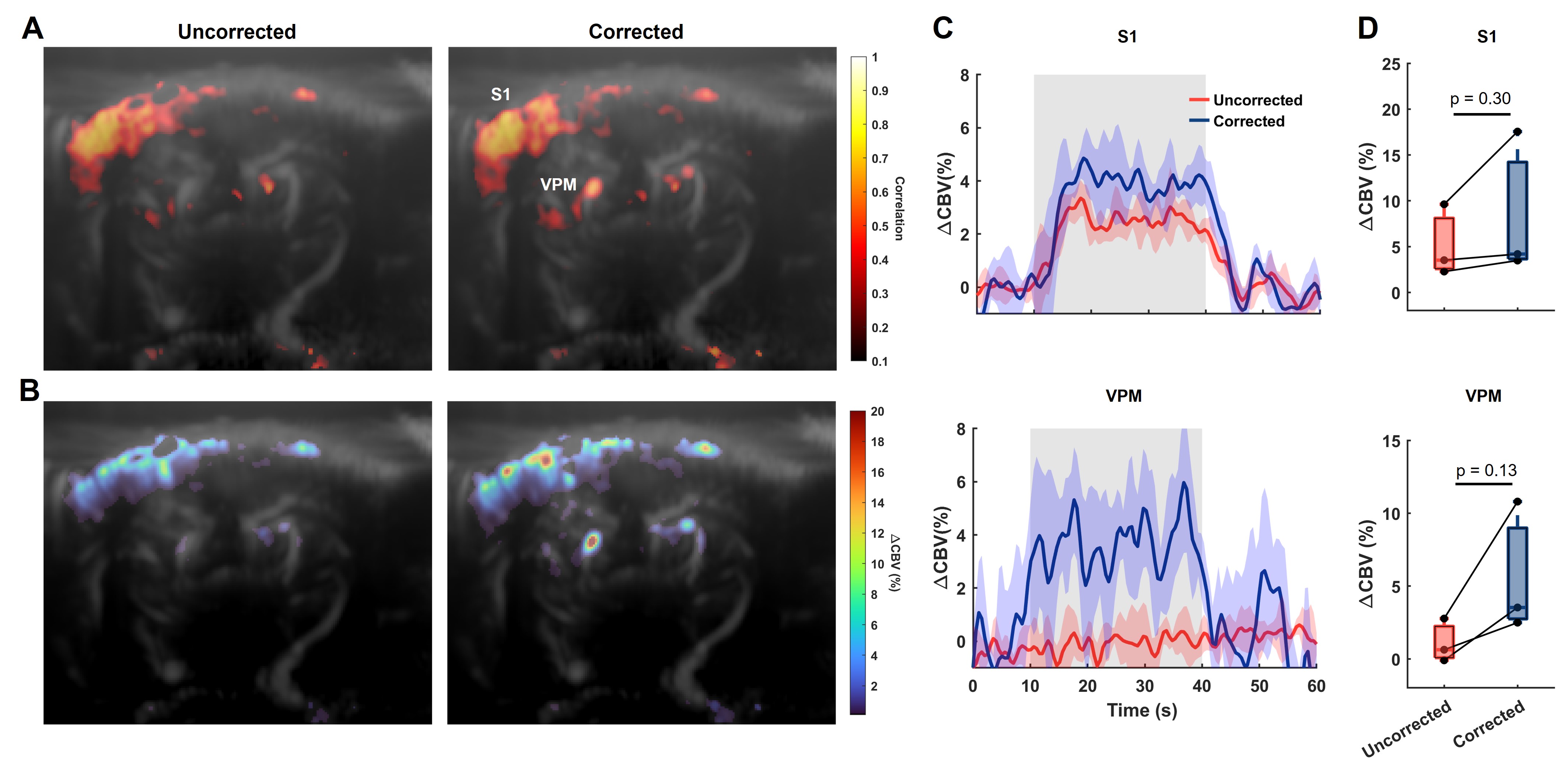}
    \caption[Aberration correction improves transcranial fUS sensitivity in mice]{Aberration correction improves transcranial fUS sensitivity in mice. (\textbf{A}) Pearson's correlation and (\textbf{B}) $\Delta$CBV maps before and after correction during whisker stimulation. (\textbf{C}) Time course changes of CBV during whisker stimulation for S1 and the VPM averaged for 5 trials. (\textbf{D}) Mean $\Delta$CBV increased from 5.1 $\pm$ 3.9 $\%$ to 8.4 $\pm$ 8.4 $\%$ after correction in S1 ($n=3$ mice, $P=$ 0.30, paired Student's \textit{t}-test) and increased from 1.1 $\pm$ 1.5 $\%$ to 5.6 $\pm$ 4.5 $\%$ in the VPM ($n=3$ mice, $P=$ 0.13, paired Student's \textit{t}-test). }
    \label{fig:fus_mice}
\end{figure*}

\begin{figure*}
    \centering
    \includegraphics[width=1\linewidth]{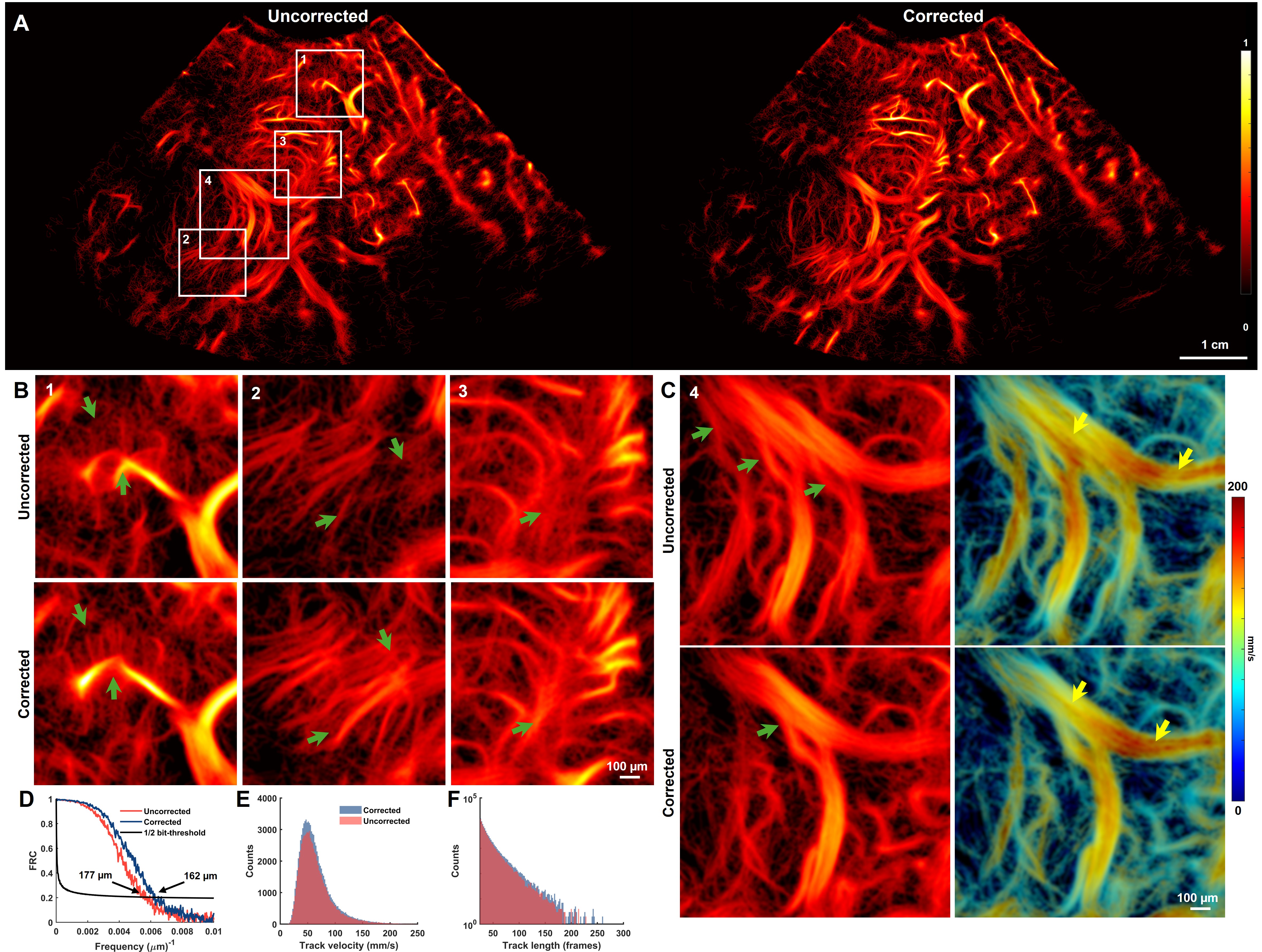}
    \caption[Transcranial ULM of the macaque brain with aberration correction]{Transcranial ULM of the macaque brain with aberration correction. (\textbf{A}) ULM maps before and after correction. (\textbf{B}) ROIs extracted from (\textbf{A}) before and after correction. (\textbf{C}) ROI extracted from (\textbf{A}) with vessel duplications and with corresponding velocity maps before and after correction. (\textbf{D}) FRC curves before and after correction. (\textbf{E}) Microbubble mean velocity distribution before and after correction ($P=$ 0.14, Kolmogorov–Smirnov test). The mean microbubble velocity was 64.6 mm/s before correction and 64.4 mm/s after correction. (\textbf{F}) Track length distribution before and after correction ($P<$ 0.0001 Kolmogorov–Smirnov test). Mean track length increased from 37.6 frames to 38.8 frames.}
    \label{fig:mcaque_ulm}
\end{figure*}

\begin{figure*}
    \centering
    \includegraphics[width=1\linewidth]{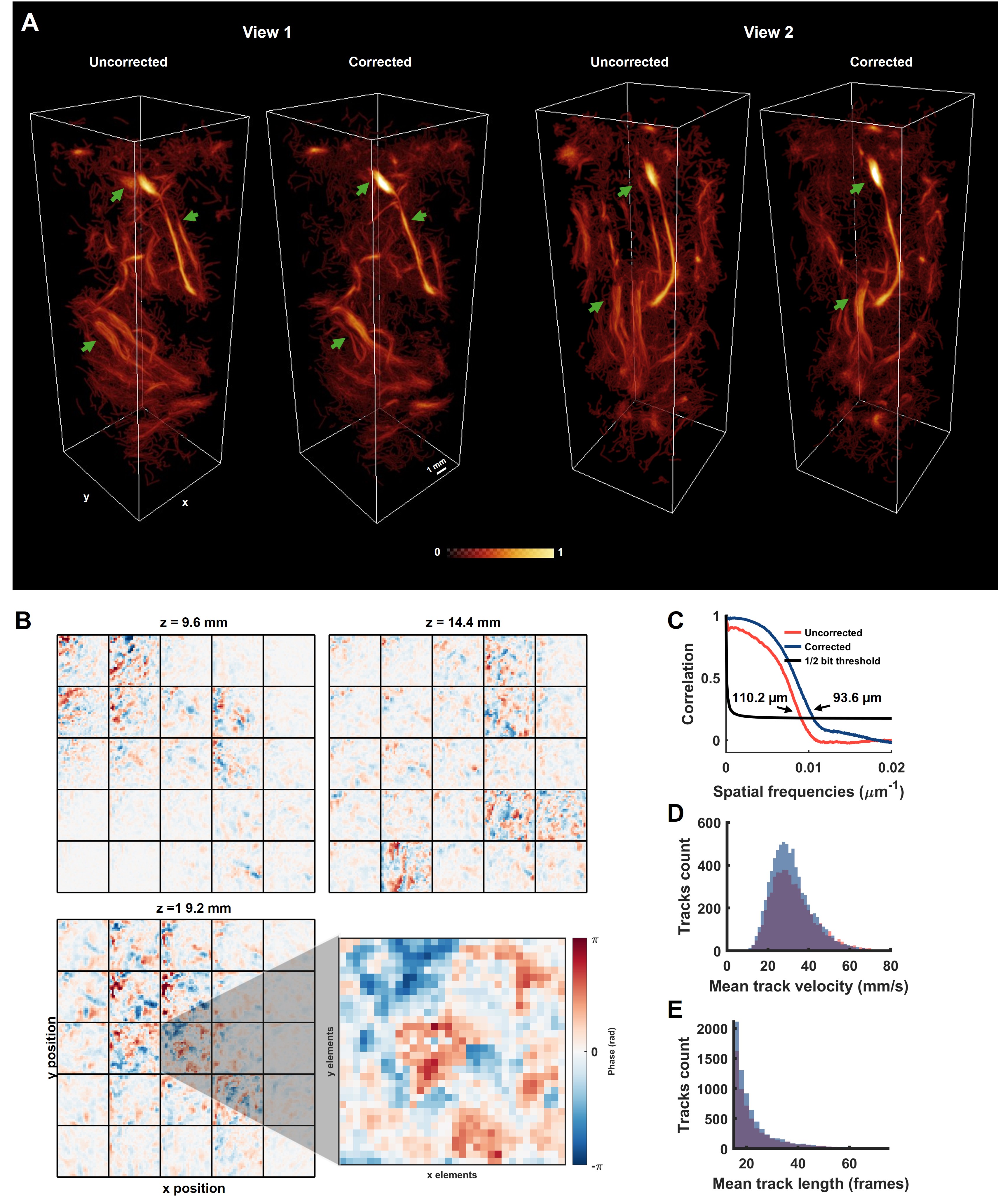}
    \caption[3D transcranial ULM of the macaque brain with aberration correction]{3D transcranial ULM of the macaque brain with aberration correction. (\textbf{A}) ULM maps before and after correction obtained in macaque 2. (\textbf{B}) Example of the depth and lateral variation of the aberration function. (\textbf{C}) FSC before and after correction showed a improvement in resolution from 110.2 to 93.6 $\mu$m. Mean track velocity decreased from 32.5 to 31.4 mm/s ($P<$ 0.0001 Kolmogorov–Smirnov test). \textbf{D}) Track mean velocity distribution. (\textbf{E}) The number of tracks of different length increased after correction, with a total increased from 5956 to 7426 tracks. }
    \label{fig:mcaque_ulm3d}
\end{figure*}

\section{Results}

\subsection{Automatic closed-loop estimation of the aberration function}

Fig. \ref{fig:methods}A illustrates the principles of transcranial aberration correction with differentiable beamforming. A spatio-temporal filter based on singular value decomposition (SVD) is performed on raw channel data to perform tissue cancellation \cite{demene2015spatiotemporal} and remove the dominant skull signal. To be able to correct for local aberration, we defined a spatially distributed aberration function parameterized with multiple isoplanatic patches covering the entire field of view. The size of the isoplanatic patches was chosen to account for the local variation in the aberration function. The aberration function is then up-sampled through a cubic interpolation to cover each pixel of the image grid, allowing for pixel-wise aberration correction with a continuous phase aberration profile. Differentiable beamforming then requires to optimize an objective function designed to improve image focusing. In particular, the magnitude of the off-diagonal terms of the angular covariance matrix $\mathbf{C}_\theta$ provides information about the level of coherence between each transmitted plane wave, which decreases when aberration is present \cite{li2017angular, bendjador2020svd}. Furthermore, the first singular vector of $\mathbf{C}_\theta$, with singular value $\sigma_1$, represents highly correlated features of the plane waves.

%and the first singular value $\sigma_1$ of $\mathbf{C}_\theta$ can be used to define the spectral norm $\vert\vert \mathbf{C}_\theta\vert\vert_2$.

Using $\sigma_1$ as an objective function (Fig. \ref{fig:methods}B), the angular coherence of the ultrasound signal can be optimized by gradient ascent. The different delays of the aberration function are automatically updated iteratively to improve focusing by moving in the direction of the gradient until reaching convergence. When differentiable beamforming is applied, the coherence between angles is increased, as well as the spatial coherence measured with the coherence factor (CF) (Fig. \ref{fig:methods}C). Therefore, differentiable beamforming allows for automatic closed-loop retrieval of the aberration function. The final aberration function can then be used to perform correction for power Doppler images (PDI) or ULM.

\subsection{Isoplanatic patch size effect on aberration correction}

Fig. \ref{fig:metrics} presents a convergence analysis of differentiable beamforming. We imaged adult mice ($n=7$, 2-3 months old) in different coronal planes using a 128-element 15-MHz linear probe. We first evaluated the impact of the number of isoplanatic patches on the capability of differentiable beamforming to correct for aberration (Fig. \ref{fig:metrics}A).  Angular coherence improved when the number of isoplanatic patches increased ($F=83.9$, $P< 0.0001$, repeated-measures one-way ANOVA), as well as the spatial coherence ($F=49.8$, $P<$ 0.0001, repeated-measures one-way ANOVA). Both metrics tended to reach a peak value when patches of size around 8$\lambda$ by 8$\lambda$ are used, where  $\lambda$ is the imaging wavelength. We also evaluated the number of iterations needed before differentiable beamforming reaches convergence. Angular and spatial coherence converged and reached a plateau before 30 iterations of gradient ascent (Fig. \ref{fig:metrics}B), which was later used as a fixed number of iterations for aberration correction. Fig \ref{fig:metrics}C shows an example of the increase in the magnitude of $\mathbf{C}_\theta$ between the initial state and the last iteration of differentiable beamforming. The angular coherence factor profile with respect to the angle index further enlightened the increase of angular coherence through differentiable beamforming. Fig \ref{fig:metrics}D then shows an example of the aberration function retrieved with differentiable beamforming after 30 iterations. Aberration profiles obtained from different patches through the different iterations are shown in  Fig \ref{fig:metrics}E. After the first iteration, the aberration profiles appeared random, but converged to stable profiles within a few steps of gradient ascent. The depth and lateral dependence of the aberration function could also be observed. Aberration profiles were mainly symmetric between the left and right hemispheres, with stronger aberration near the skull (Fig. \ref{fig:metrics}F, $F=3.65$, $P<$ 0.01, two-way ANOVA with post-hoc \textit{t}-test and Bonferroni correction).

\subsection{Transcranial Doppler and CEUS Doppler of the mouse brain}

Next, we evaluated the capability of differentiable beamforming to retrieve the aberration function in the presence or absence of microbubbles. Skull-induced aberration could be corrected in the absence of microbubble (Fig \ref{fig:pdi}A-D). In 2-3 months old mice, the contrast-to-noise ratio (CNR) improved from 23.1 $\pm$ 5.7 dB  to 24.3 $\pm$ 6.1 dB after correction ($n=7$, $P  < $ 0.05, two-tailed paired Student's \textit{t}-test), and the signal-to-noise ratio (SNR) improved from 17.8 $\pm$ 5.6 dB to 19.0 $\pm$ 6.2 dB ($n=7$, $P  < $ 0.05, two-tailed paired Student's \textit{t}-test). Improvement in resolution reduced selected vessel widths from 202.4 $\pm$ 50.2 $\mu$m to 162.6 $\pm$ 27.8 $\mu$m ($n=7$ mice, $P  < $ 0.05, two-tailed paired Student's \textit{t}-test) and tended to increase vessel intensity ($n=7$, $P  = $ 0.17, two-tailed paired Student's \textit{t}-test). However, because the skull remains highly attenuating, strong shadowed area persisted even after aberration correction, particularly in older animals (Fig. \ref{fig:pdi}B-C). In 6-7 months old mice, the CNR improved from 19.5 $\pm$ 3.9 dB to 20.6 $\pm$ 4.3 dB after correction ($n=3$, $P  = $ 0.09, two-tailed paired Student's \textit{t}-test) and the SNR improved from 14.5 $\pm$ 5.0 dB to 15.6 $\pm$ 5.4 dB ($n=3$, $P  = $ 0.08, two-tailed paired Student's \textit{t}-test). The widths of selected vessels decreased from 178.2 $\pm$ 20.7 $\mu$m to 144.1 $\pm$ 28.7 $\mu$m ($n=3$ mice, vessels, $P  = $ 0.14, two-tailed paired Student's \textit{t}-test).

The use of microbubble through bolus injection improved transcranial vascular imaging by compensating for skull attenuation and recovered vessels in previously shadowed areas (Fig \ref{fig:pdi}E-F). In 2-3 months old mice, the CNR improved from 35.2 $\pm$ 4.8 dB to 37.8 $\pm$ 4.9 dB ($n=7$, $P  < $ 0.001, two-tailed paired Student's \textit{t}-test) and the SNR from 30.5 $\pm$ 5.4 dB to 32.9 $\pm$ 5.2 dB ($n=7$, $P  < $ 0.001, two-tailed paired Student's \textit{t}-test) while in 6-7 months old mice, the CNR improved from 35.7 $\pm$ 5.5 dB to 38.8 $\pm$ 3.6 dB ($n=4$, $P  < $ 0.05, two-tailed paired Student's \textit{t}-test) and the SNR from 28.8 $\pm$ 5.7 dB to 32.0 $\pm$ 4.7 dB ($n=4$, $P  < $ 0.05, two-tailed paired Student's \textit{t}-test). In younger mice, the selected vessel intensity significantly increased ($n=7$, $P  < $ 0.05, two-tailed paired Student's \textit{t}-test) and their width decreased from 230.5 $\pm$ 50.5 $\mu$m to 167.8 $\pm$ 38.7 $\mu$m ($n=7$ mice, $P  < $ 0.01  two-tailed paired Student's \textit{t}-test). In older mice, vessels also tended to be sharper after correction ($n=4$, $P  = $ 0.10, two-tailed paired Student's \textit{t}-test), with their width decreasing from 287.9 $\pm$ 68.4 $\mu$m to 216.6 $\pm$ 73.8 $\mu$m ($n=4$ mice, $P  < $ 0.05, two-tailed paired Student's \textit{t}-test).

%\subsection{Microbubble concentration effect on aberration correction}

%To evaluate the effect of microbubble concentration on differentiable beamforming capability to retrieve the aberration function, we applied aberration correction using ultrasound acquisition at different time points of the bolus injection. Our results showed that differentiable beamforming was able to perform a similar aberration correction independently of the microbubble concentration. Figure \ref{fig:concentration}a shows that CNR and SNR increased with microbubble concentration in young (CNR: $F=13.2$, $P  < $ 0.0001, SNR: $F=9.4$, $P  < $ 0.0001, repeated-measures two-way ANOVA) and old mice (CNR: $F=9.2$, $P  < $ 0.001,  SNR: $F=7.2$, $P < $ 0.001, repeated-measures two-way ANOVA). The increase in CNR and SNR after aberration correction remained significant at different microbubble concentrations for young (CNR:$F=26.0$, $P  < $ 0.01, SNR: $F=24.5$, $P  < $ 0.01, repeated-measures two-way ANOVA) and old mice (CNR: $F=12.0$, $P  < $ 0.05, SNR: $F=10.2$, $P  < $ 0.05, repeated-measures two-way ANOVA).

To evaluate the effect of microbubble concentration on differentiable beamforming capability to retrieve the aberration function, we applied aberration correction using ultrasound acquisition at different time points of the bolus injection. Our results showed that differentiable beamforming was able to perform a similar aberration correction independently of the microbubble concentration. Figure \ref{fig:concentration}a shows that CNR and SNR increased with microbubble concentration in young and old mice. The increase in CNR and SNR after aberration correction remained significant for the different microbubble concentrations. The vessel width improved regardless of the microbubble concentration in young but did not reach significance for old mice (Fig. \ref{fig:concentration}B). There was no significant effect of microbubble concentrations on vessel width and the vessel intensity tended to be higher after correction. Spatial and angular coherence also increased for the different microbubble concentrations (Fig \ref{fig:concentration}C). The mean standard deviation of the aberration function was 0.20 radians or less for the different concentrations analyzed (Fig \ref{fig:concentration}D). The aberration functions also showed on average a correlation coefficient higher than 0.85 for young mice and higher than 0.60 for old mice (Fig. \ref{fig:concentration}E), further confirming the robustness of differentiable beamforming to variation of microbubble concentrations.

%The combined effect of the microbubble concentration and aberration correction was not significant for young mice (CNR: SNR: $P=0.27$), but significant for older mice (CNR: $P < 0.05 $, SNR: $P < 0.05 $), which could be explained by the lower signals at low concentrations when the skull is thicker. 

%The vessel width improved regardless of the microbubble concentration in young (Fig \ref{fig:concentration}b, $P  < $ \hl{0.05}) but did not reach significance for old mice ($P  = $ \hl{0.27}). There was no significant effect of microbubble concentrations on vessel width (young mice: $P  = $ \hl{0.97}, old mice: $P  = $ \hl{0.61})) or combined effect on vessel width improvement after correction (young mice: $P  = $ \hl{0.23}, old mice: $P  = $ \hl{0.56}). Vessel intensity across microbubble concentration tended to be higher after correction for young ($P  = $ \hl{0.051}) and old mice ($P = $\hl{0.30}).

%The vessel width improved regardless of the microbubble concentration in young (Fig. \ref{fig:concentration}b, $F=10.6$, $P  < $ 0.05, repeated-measures two-way ANOVA) but did not reach significance for old mice ($F=1.8$, $P  = $ 0.27, repeated-measures two-way ANOVA). There was no significant effect of microbubble concentrations on vessel width (young mice: $F=0.2$, $P  = $ 0.97, old mice: $F=0.8$, $P  = $ 0.61, repeated-measures two-way ANOVA). The vessel intensity tended to be higher after correction (young mice: $F=5.9$, $P  = $ 0.051, old mice: $F=1.6$, $P = $ 0.30, repeated-measures two-way ANOVA).

\subsection{Transcranial ULM imaging of the mouse brain}

Because ULM relies on the precise localization of microbubbles, we evaluated the effect of aberration correction on the microbubble detection performance. The presence of skull-induced aberration caused distortion in microbubble shapes marked by blurred focal spots (Fig. \ref{fig:supp_microbubbles}A). The use of differentiable beamforming improved the microbubble shapes, which were sharper and more symmetric after correction (Fig. \ref{fig:supp_microbubbles}B). The number of detected microbubbles also increased, as shown by the detection rate over time in Fig. \ref{fig:supp_microbubbles}C.

We then performed transcranial ULM reconstructions of the mouse brain vasculature using differentiable beamforming. For 2-3 months old mice, the presence of skull-induced aberration decreased localization and tracking performance, marked by regions of the brain where vessels appeared blurred, missing, or disconnected (Fig. \ref{fig:ULM_mice}A-B). Not only did the vessels appear sharper after correction, additional vessels could be observed, and previously disjoint vessels were properly connected.

Measurements with the Fourier ring correlation (FRC)-based method \cite{hingot2021measuring} using the half-bit threshold showed that the ULM effective resolution improved from 23.8 $\pm$ 5.4 $\mu$m to 18.4 $\pm$ 3.0 $\mu$m after correction, as shown by the FRC curves in Fig. \ref{fig:ULM_mice}C. The improvement in resolution for the different animals was statistically significant ($ n=7, P  < $ 0.01, two-tailed paired Student's \textit{t}-test). To further evaluate the performance of ULM, we computed the saturation curves, which by counting the number of illuminated pixels during reconstruction, show the capability of ULM to detect additional vascular structures \cite{heiles2022performance}. The results show that aberration correction increased the saturation area under the curve (AUC) from 2.10 $\pm$ $0.26\times10^4$ to  2.21 $\pm$ $0.25\times10^4$ (Fig. \ref{fig:ULM_mice}D, $n=7$, $P  < $ 0.05, two-tailed paired Student's \textit{t}-test) and increased the number of tracks from 2.84 $\pm$ $0.88\times10^5$ to  3.90 $\pm$ 0.96 $\times10^5$ (Fig. \ref{fig:ULM_mice}E, $ n=7, P  < $ 0.001, two-tailed paired Student's \textit{t}-test), supporting the presence of additional microvessels.

Our results showed that differentiable beamforming was also effective to correct for stronger aberration due to thicker skull in older mice that were 6-7 months old (Fig. \ref{fig:ULM_mice}F-J). The reconstructed brain vasculature suffered from strips artifact caused by the skull shadowing \cite{errico2015ultrafast, renaudin2022functional}, which are characterized by the presence of dark regions where the vasculature is absent. Part of the vasculature in regions displaying the strong shadowing artifacts could be recovered after aberration correction. The resolution significantly improved from 28.0 $\pm$ 3.5 $\mu$m to 23.6 $\pm$ 3.7 $\mu$m (Fig. \ref{fig:ULM_mice}H,  $ n=4, P < $ 0.05, two-tailed paired Student's \textit{t}-test). Although saturation did not increase (Fig. \ref{fig:ULM_mice}I, $ n=4, P  = $ 0.15, two-tailed paired Student's \textit{t}-test), the number of tracks increased from 4.32 $\pm$ 0.40 $\times10^5$ to 4.78 $\pm$ 0.60 $\times10^5$ (Fig. \ref{fig:ULM_mice}J, $ n=4, P  < $ 0.05, two-tailed paired Student's \textit{t}-test).

\subsection{Effect of aberration on hemodynamic quantification in ULM}

Beyond vascular structure maps, ULM also allows to establish hemodynamic quantification of the brain microvasculature. Hence, ULM could facilitate the development of new biomarkers based on the detection of hemodynamic changes that occur in cerebrovascular diseases, as shown in previous studies \cite{demene2021transcranial, lowerison2022aging, lowerison2024super, lin2024super}. We aimed to evaluate how skull-induced aberration affects hemodynamics in ULM (Fig. \ref{fig:ulm_hemodynamics}). An example of a reconstructed directional velocity map of the mouse brain after aberration correction is shown in Fig. \ref{fig:ulm_hemodynamics}A. Vector fields extracted from the ULM velocity maps can reveal additional information on vascular flow dynamics of the brain, such as direction and type of flow along vessels (Fig. \ref{fig:ulm_hemodynamics}C). In aberrated areas, vascular flow can appear turbulent, which is caused by changes in flow direction and could be misinterpreted as pathological conditions \cite{demene2021transcranial}. Laminar flow similar to normal physiological conditions can be recovered using differentiable beamforming, with microbubbles following a smooth path along the vessels, showing that aberration correction is crucial for reliable blood flow characterization in the brain. Furthermore, velocity profile views in selected vessels showed that differentiable beamforming enables retrieving parabolic blood flow profiles along the different vascular trees that are consistent with physiological conditions, while in the presence of aberration, these profiles appear to be more disrupted (Fig. \ref{fig:ulm_hemodynamics}D). 

We then analyzed the microbubble flow speed distributions throughout the brain in the presence and absence of aberration. After correction, the microbubble velocity distribution decreased (Fig. \ref{fig:ulm_hemodynamics}E, $P<$ 0.0001 Kolmogorov–Smirnov test), which is consistent with a larger number of smaller vessels reconstructed, as well as less spurious velocity profiles. The difference in microbubble velocity was observed in multiple animals, decreasing from 14.1 $\pm$ 2.4 to 11.3 $\pm$ 2.0 in young mice ($n=7$, $P  < $ 0.001, two-tailed paired Student's \textit{t}-test) and decreasing from 14.1 $\pm$ 0.8 mm/s to 11.7 $\pm$ 0.3 mm/s in old mice ($n=4$, $P  < $ 0.05, two-tailed paired Student's \textit{t}-test). Detected microbubbles have also longer tracks (Fig. \ref{fig:ulm_hemodynamics}G, $P<$ 0.0001 Kolmogorov–Smirnov test), increasing from an average length of 37.6 $\pm$ 2.4 to from 43.7 $\pm$ 3.4 in young mice ($n=7$, $P  < $ 0.0001, two-tailed paired Student's \textit{t}-test) and increasing from 25.0 $\pm$ 1.9 frames to 28.5 $\pm$ 1.5 frames in old mice ($n=4$, $P  < $ 0.01, two-tailed paired Student's \textit{t}-test), as shown in Fig. \ref{fig:ulm_hemodynamics}H. Fig. \ref{fig:ulm_hemodynamics}e shows the increase in the number of tracks for different track lengths. Hence, after aberration correction, microbubbles can be followed at larger distances as they flow through the vasculature. Longer tracks allow not only for improved vessel delineation but also more reliable hemodynamic quantification since velocity can be continuously computed on a longer distance.

%Quantitative assessment of velocity profiles across the vessel sections showed that significant speed differences are better detected after aberration correction ($P<$ \hl{XX}, ANOVA with post-hoc \textit{t}-test and Bonferroni correction). Using differentiable beamforming, velocity difference within vessels could be measured using bins less than \hl{XX} $\mu$m wide.

\subsection{Transcranial functional imaging in mice}

fUS can image brain-wide neurovascular activities with high spatiotemporal resolution when compared to other modalities, such as functional magnetic resonance imaging (fMRI). Transcranial fUS imaging remains challenging due to skull-induced aberration and attenuation that reduce Doppler sensitivity \cite{errico2016transcranial, tiran2017transcranial, vert2025transcranial}. As a result, most fUS studies are performed through a craniotomy of a thinned-skull cranial window \cite{mace2011functional, osmanski2014functional, brunner2020platform}. 

To evaluate how differentiable beamforming can improve fUS sensitivity, we performed transcranial fUS imaging in three additional mice (8-9 weeks old) during whisker stimulation using a 15-MHz probe placed over the primary somatosensory cortex (S1). To detect activated brain areas during stimulation, both Pearson's correlation coefficient and the relative cerebral blood volume ($\Delta$CBV) were computed. Fig \ref{fig:fus_mice}A-B present the activation maps before and after aberration correction. Activation across the barrel cortex in S1 could be observed during whisker stimulation on both the correlation and $\Delta$CBV maps. However, activation in the ventral posteromedial nucleus (VPM) of the thalamus could only be observed after aberration correction. Fig \ref{fig:fus_mice}C shows that the time course of $\Delta$CBV through the stimulation presented a larger stimulus-evoked increase in the hemodynamic response after aberration correction. Fig \ref{fig:fus_mice}D shows that the mean $\Delta$CBV amplitude during stimulation in both S1 ($n=3$, $P=$ 0.30, paired Student's \textit{t}-test) and VPM ($n=3$, $P=$ 0.13, paired Student's \textit{t}-test) tended to increase after correction.

%Our results show that aberration correction is necessary to correct for anatomical locations of vascular structures and corresponding cerebral activity

\subsection{Nonhuman primate transcranial brain imaging}

To demonstrate that differentiable beamforming can be generalized to different imaging configurations and used in a setting similar to clinical transcranial imaging, we next translate our experiments to NHP using an adult Rhesus macaque. To cover a larger field of view needed to image the macaque brain, we used a phased array probe with divergent wave transmissions which are also adapted for transcranial imaging in humans \cite{demene2021transcranial}. We also imaged with a lower frequency to compensate for the higher attenuation of the macaque skull, the thickness of which is comparable to the human temporal bone \cite{adams2007biocompatible, kwon2006thickness}. Fig. \ref{fig:mcaque_ulm} shows transcranial ULM maps obtained with a 64-element probe driven at 2.5 MHz and placed above the motor cortex. Before correction, major aberration artifacts, such as vessel duplications, can be observed in deep seating vessels around the circle of Willis, as well as disconnected vessels, particularly in the lenticulostriate arteries arising from the middle cerebral artery (MCA).

Differentiable beamforming improved ULM reconstruction, removing the vessel duplications around the circle of Willis, and allowing the construction of sharper and better connected vessels (see Fig. \ref{fig:mcaque_ulm}B). The microvascular architecture of the cortical vessel branches arising from the pial vessels (see Fig. \ref{fig:mcaque_ulm}B, panel 1) and of the lenticulostriate arteries (see Fig. \ref{fig:mcaque_ulm}B, panel 3) could be clearly observed on the ULM map after correction. Aberration correction also improved velocity mapping, as shown in Fig. \ref{fig:mcaque_ulm}C. Velocities profile along vessels are more continuous, as shown by the velocity map, notably in the circle of Willis area. The FRC showed that the resolution improved from 177 $\mu$m to 162 $\mu$m (Fig. \ref{fig:mcaque_ulm}D). After correction microbubble velocity distribution also tended to be lower (Fig. \ref{fig:mcaque_ulm}E, $P=$ 0.13 Kolmogorov–Smirnov test) and the track lengths increased (Fig. \ref{fig:mcaque_ulm}F, $P<$ 0.0001, Kolmogorov–Smirnov test).

%Parabolic flow profiles could be retrieved after correction, consistent with normal physiological conditions.

\subsection{3D transcranial ULM in nonhuman primates}

Although 2D imaging can capture the vasculature and hemodynamics of the brain, it still suffers from many limitations, such as the need for plane selection and bias caused by projection of 3D structures into the 2D plane. To overcome these limitations, we then demonstrated that differentiable beamforming can be extended to 3D imaging. To establish the translational potential for clinical application, we again used the Rhesus macaque as a model, imaging through the intact skull and skin using a 3-MHz multiplexed matrix-array probe. We defined a spatial aberration function that captures both lateral and depth variations of aberration caused by the skull.

In Fig \ref{fig:mcaque_ulm3d}A, major artifacts induced by the skull, such as multiple vessel duplications, are removed with differentiable beamforming. Fig. \ref{fig:mcaque_ulm3d}B shows the lateral variation of the aberration function obtained at three different depths and computed using only 5 transmit angles. The Fourier shell correlation (FSC) showed that the resolution improved from 110.2 to 93.6 $\mu$m (Fig \ref{fig:mcaque_ulm3d}C). After correction, the mean track velocity also decreased from 32.5 to 31.4 mm/s (Fig \ref{fig:mcaque_ulm3d}D, $P<$ 0.0001 Kolmogorov–Smirnov test) and the number of detected tracks increased (Fig \ref{fig:mcaque_ulm3d}E). Two additional transcranial acquisitions in macaques are shown in Fig. \ref{fig:supp_3d}. Although aberration was less severe in these cases, the use of differentiable beamforming sill improved ULM, allowing for sharper and brighter vessel reconstruction.

\section{Discussion}

In this work, we presented a differentiable beamforming framework taking advantage of open-source deep learning libraries to perform automatic gradient computation for aberration correction to image through the skull. By exploiting a spatially-distributed parametrization of the aberration function, we optimized in a closed-loop manner the ultrasound signal angular coherence and were able to achieve robust transcranial imaging of the brain hemodynamics both for power Doppler images and super-resolution ULM.

Aberration correction increased the microbubble detection rate but also increased the microbubble track lengths. Smaller vessels are better reconstructed and previously disjoint or duplicated vessels are better connected. ULM resolution improved in younger mice and in older mice. Differentiable beamforming was also effective to image transcranially in a NHP using a phased array in settings similar to clinical applications. We also showed that differentiable beamforming could be extended to 3D transcranial ULM of the macaque brain by optimizing the coherence between only 5 transmit angles, capturing not only depth but also lateral variation of the aberration. The macaque skull thickness is comparable to the human temporal bone often used as an imaging window in ultrasound, suggesting that our approach could also be effective to correct for aberration in humans.

Beyond improved vascular anatomical maps of the brain microvasculature, this work highlights the importance of performing aberration correction for reliable hemodynamic quantification in ULM, as shown by previous work \cite{xing2024phase, xing2025inverse}. We showed that correcting for skull aberration allowed recovery of normal physiological parabolic flow profiles in vessels as well as vector fields consistent with laminar flow. This could be particularly relevant in the context of transcranial longitudinal studies \cite{mccall2023longitudinal} where growing skull thickness could hinder ULM reconstruction at older ages and could induce bias in blood flow estimations and dynamics. Clinical applications would also benefit from the proposed correction method, with a better diagnosis of cerebrovascular diseases \cite{demene2021transcranial}.

Aberration correction could also improve the sensitivity of transcranial fUS imaging of brain activity. We showed that differentiable beamforming allowed for more accurate delineation of cerebrovascular activity. Such improvements in image resolution allowing for more precise anatomical locations of vascular structures would help detect single-whisker events using fUS \cite{mace2011functional} and study the spatial specificity of neurovascular coupling \cite{martineau2024distal} transcranially and across brain areas. Current clinical applications of fUS imaging are also limited to newborns \cite{demene2017functional, baranger2021bedside}, intraoperative settings \cite{imbault2017intraoperative, soloukey2020functional} or through acoustically transparent cranial implants \cite{rabut2024functional, soloukey2025mobile}. By improving ultrasound sensitivity and correcting for vascular distortions, transcranial aberration correction constitutes a first step toward non-invasive fUS imaging of the human brain.

Our work follows recent developments on differentiable beamforming \cite{simson2023differentiable, heriard2025path}, which were used to perform adaptive focusing through speed of sound mapping. The main advantage of our proposed framework and aberration parameterization is that it does not require the prior acquisition of a separate synthetic aperture or full plane wave dataset to perform correction and can be performed in principle on any ULM dataset. Differentiable beamforming can be performed in both the absence and in the presence of microbubbles. Hence, differentiable beamforming does not require the need for guide stars \cite{demene2021transcranial, robin2023vivo, xing2024phase, xing2025inverse} and could in principle be applied to any microbubble concentration conditions.

%However, we did not evaluate the impact of the number of plane waves on aberration correction performance, but having a higher number of transmits should yield more angular frequency components and a more complete $\mathbf{C}_\theta$ for optimization

%We used the spatial CF along the ACM as an objective function for aberration correction. Biological tissues are characterized by speckle, i.e., a random distribution of scatterers that can be modeled as an incoherent source, which is known from the van Cittert-Zernike theorem to show only partial coherence along the aperture \cite{mallart1991van}. Hence, low CF could also be caused by speckle decorrelation across the aperture and not only by the presence of aberration \cite{simson2024ultrasound}, and speckle jitter could be overfitted during gradient ascent. However, this could be partially alleviated when microbubbles are used, which act as coherent point source scatterers and should hence display higher coherence across the aperture, and by the use of the ACM, which evaluates the coherence between plane waves across all pixels. 

Although we showed that differentiable beamforming can be used to perform aberration correction for transcranial imaging in rodents and NHPs, this study is not without limitations. Other works on differentiable beamforming investigated alternative objective or loss functions that may prove to be more reliable for aberration correction. The common midpoint phase error was recently introduced to optimize speed of sound map with better performance than other metrics such as the CF \cite{simson2023differentiable, simson2024ultrasound}. However, the use of full synthetic aperture data with lower frame rate limits (around 100 Hz) could reduce the capability of ultrasound to image blood flow in the brain. Standard ULM acquisitions are usually composed of only a few plane waves, which also prevents the use of decoding strategies to recover the complete synthetic aperture dataset \cite{bottenus2017recovery}. Hence, using the angular covariance matrix, we were still able to exploit the coherence of backscattered echoes between different transmits using a limited number of plane waves. The proposed correction method relies on the definition of isoplanatic patches. There could be a size limit of the isoplanatic patch that can be solved with the differentiable beamforming and the optimal way to determine the ideal number of patches could be further investigated. Although the patch size was determined heuristically in this study, it remained consistent between multiple animals even when imaged at different coronal planes. The patch size was also comparable to other studies, even if the experimental settings were different \cite{bendjador2020svd, lambert2022bultrasound}. 

%Nevertheless, a different aberration model, such as the recently path-based parametrization \cite{heriard2025path} could be more computationally efficient to correct for skull-induced aberration than the model we proposed in this paper.

On a more general note, this work strengthens the importance of physics-informed differentiable algorithms to solve physics-based problems. Such frameworks can be used to solve optimization problems on a single dataset without prior training requirement and can be easily adapted for different emission schemes such as plane waves or divergent waves and from 2D to 3D imaging. Differentiable beamforming could facilitate the large-scale deployment of aberration correction strategies beyond the initial proof-of-concept settings. Finally, automatic differentiable computing can be seen as a general framework for optimization that can be applied to any parameters of interest, enabling novel applications in medical imaging.

\section{Methods}

\subsection{Differentiable beamforming with spatially-distributed aberration parametrization}

A differentiable in-phase quadrature (IQ) delay-and-sum (DAS) beamformer was implemented in Python (version 3.12.8) using the PyTorch machine learning library (version 2.5.1), which provides both complex number automatic differentiation and graphics processing unit (GPU) acceleration frameworks. DAS beamforming is widely used for image formation in ultrasound imaging due to its simple implementation and fast computation. Using DAS, the pixel $s(\mathbf{r})$ of the image can be reconstructed by computing the combined forward and backward travel time $\tau_{n,i}(\mathbf{r})$ for each receiving element $n$ and transmission $i$ to the position $\mathbf{r}=(x,z)$, and then by the coherent summation of the corresponding receive aperture signal $s_{n,i}(\tau_{n,i}(\mathbf{r}))$, such as
\begin{align}
    s(\mathbf{r}) = \sum_{i=1}^{N_{tx}}\sum_{n=1}^{N_{ele}}s_{n,i}(\tau_{n,i}(\mathbf{r}))e^{i2\pi f_c \tau_{n,i}(\mathbf{r})},
\end{align}

where $N_{ele}$ represents the number of elements of the ultrasonic probe, $N_{tx}$ the number of transmissions, and $f_c$ the central frequency. Because the received signal $s_{n,i}(.)$ here is assumed to be IQ demodulated, a phase rotation must be applied before summation. A cubic interpolation was used to evaluate the $s_{n,i}(\tau_{n,i}(\mathbf{r}))$ terms before the delay summation. 

DAS generally assumes a globally uniform speed of sound in the medium, allowing the use of simple geometric considerations for computing the different time-of-flights (TOFs). For plane wave imaging using a linear probe with transmit angle $\theta_i$, the different travel times can be computed as 
\begin{align}
    \tau_{n,i}(\mathbf{r})= \frac{z\cos\theta_i+x\sin\theta_i+\sqrt{z^2+(x-x_n)^2}}{c},
\end{align}

where $c$ is the speed of sound and $x_n$ the position of the transducer. Aberration occurs when heterogeneities are present, as is the case with the skull for brain imaging, leading to a mismatch between the assumed and real TOFs. Hence, aberration correction can be performed by compensating the theoretical geometrical delay by including an additional time delay $u_n(\mathbf{r})$ in order to account for the travel time difference
\begin{align}
   \tau_{n,i}^{corr}(\mathbf{r})= \tau_{n,i}(\mathbf{r})+ u_n(\mathbf{r}),
\end{align}

which can then be included into the beamformer
\begin{align}
        s^{corr}(\mathbf{r};\mathbf{u}) = \sum_{i=1}^{N_{tx}}\sum_{n=1}^{N_{ele}}s_{n,i}( \tau_{n,i}^{corr}(\mathbf{r}))e^{i2\pi f_c \tau_{n,i}^{corr}(\mathbf{r})}.
\end{align}

In the next sections, we will refer to the matrix $\mathbf{u}$ as the aberration function, which contains all $u_n(\mathbf{r})$ delays.

\subsection{Focus quality as an objective function}

Aberration correction can be broadly formulated as an optimization problem aiming to maximize an objective function of $\mathbf{u}$ that is designed to measure the focus quality of the image. One such criteria is the coherence between each plane wave transmission \cite{li2017angular}. Considering the beamformed images for each transmit angle, the ultrasound data can be organized into the matrix $S(\mathbf{r},\theta;\mathbf{u})$ of dimension $N_xN_z\times N_{tx}$, with $N_x$ the number of pixels in the $x$ dimension and  $N_z$ the number of pixels in the $z$ dimension. The angular covariance matrix $\mathbf{C}_\theta$ \cite{bendjador2020svd} can then be defined as
\begin{align}
\mathbf{C}_\theta(\mathbf{u}) = {\tilde{S}(\mathbf{r},\theta;\mathbf{u})}^{\dagger} \cdot \tilde{S}(\mathbf{r},\theta;\mathbf{u}),    
\end{align}

where the symbol ${\dagger}$ stands for the transpose conjugate, $\sim$ the column-wise vector normalization and $\cdot$ the  matrix multiplication. The off-diagonal terms of $\mathbf{C}_\theta(\mathbf{u})$ can be used to quantify focus quality. In fact, the profile along the anti-diagonal can be interpreted as a generalization of the Van Cittert Zernike theorem \cite{mallart1991van} to the angular domain. Hence, maximizing the magnitude of the angular covariance matrix is equivalent to optimizing the focus quality. Notably, the first singular value of $\mathbf{C}_\theta(\mathbf{u})$ is related to the level of angular coherence. Hence, to optimize the angular coherence, we defined the objective function as the spectral norm of $\mathbf{C}_\theta(\mathbf{u}) $
\begin{align}
    f(\mathbf{u}) = \vert\vert\mathbf{C}\vert\vert_2=  \sigma_{1},
\end{align}
where $\sigma_{1}$ represents the first singular value. Furthermore, an optional regularization term $ R(\mathbf{u})$  can be included as a penalty in the optimization problem to facilitate convergence and to constrain the solution space. Then performing aberration correction can be rewritten as finding $\hat{\mathbf{u}}$, the solution that maximizes the regularized objective function
\begin{align}
    \hat{\mathbf{u}} = \text{argmax}_{\mathbf{u}} \{f(\mathbf{u}) - \alpha R(\mathbf{u})\}.
\end{align}

\subsection{Regularization}

%Although squaring the L1 norm is not common, we found that this strategy performed well when the number of isoplanatic patches increased.
To avoid overfitting, a L1 regularization was used, adding a penalty to the objective function that is proportional to the L1-norm of $\mathbf{u}_{}$ and limiting the effect of noisy signals. In order to retrieve smooth solutions of $\mathbf{u}_{}$ with continuous aberration profiles, the second derivative of $\mathbf{u}_{}$ along the element dimension was also used as regularization term, as previously demonstrated in \cite{xing2025inverse}. Because $\mathbf{u}_{}$ also varies spatially, an additional regularization term based on the Laplacian operator was used to obtain smooth spatial solutions. $R(\mathbf{u})$ was set as
\begin{align}
 R(\mathbf{u}) = \alpha_1||\nabla_{(x,z)}^2\mathbf{u}||_2^2 + \alpha_2||\mathbf{u}''||_2^2 + \alpha_3||\mathbf{u}||_1,
\end{align}

where $\nabla_{(x,z)}^2$ was applied using the convolution with the nine-point stencils discrete Laplacian operator
\begin{align}
    \mathbf{D}_{(x,z)}^2 = \frac{1}{4}
    \begin{bmatrix}
    1 &2 & 1\\
    2 & -12  & 2\\
    1 & 2 & 1
    \end{bmatrix},
\end{align}
and where $\alpha_1$, $\alpha_2$, and $\alpha_3$ represent a regularization hyperparameters.

\iffalse
such as
\begin{align}
    \hat{\mathbf{u}} = \text{argmax}_{\mathbf{u}} \{f(\mathbf{u}) - \alpha ||\mathbf{D_2}{\mathbf{u}} || ^2\},
\end{align}
where $\mathbf{D}_2$ represents the second derivative matrix applied along the aperture dimension.
\fi

\subsection{Gradient ascent optimization}

DAS, being composed of linear differentiable operators, can therefore be made differentiable. Automatic differentiation frameworks such as PyTorch and Google JAX allow us to compute differentiation with respect to any parameters of interest without the explicit need of the analytical expression of the gradient. Taking advantage of such frameworks, the optimization problem can be solved via gradient ascent. 

More precisely, stepping in the direction of the gradient of a function allows us to move in a trajectory that maximizes its value. Starting from an initial guess $ \mathbf{u}_{0}$, the aberration function $ \mathbf{u}_{}$ can hence be iteratively updated by small steps in order to increase the value of $f(\mathbf{u})$ such as 
\begin{align}
    \mathbf{u}^{(j+1)} =     \mathbf{u}^{(j)} + \gamma\nabla_{\mathbf{u}} f(\mathbf{u}^{(j)}),
\end{align}

where $\nabla_{\mathbf{u}}$ represents the gradient with respect to $\mathbf{u}$ which is computed by backpropagation. The hyperparameter $\gamma$ represents the learning rate and can be fine-tuned by using different optimizer algorithms such as the Adam optimizer commonly used in deep learning frameworks. Gradient ascent can be applied for a fixed number of iterations or until reaching a convergence criterion, such as 
\begin{align}
\lVert f(\mathbf{u}^{(j+1)}) - f(\mathbf{u}^{(j)})\rVert < \epsilon,    
\end{align}

where $\epsilon$ represents the tolerance value.

\subsection{isoplanaticity and multi-patches parametrization of the aberration function}

The aberration function $\mathbf{u}$ is composed of delay adjustments accounting for the heterogeneity of the imaging medium. Assuming that the aberration profile is independent of the pixel location, then $\mathbf{u}$ takes the form of a single set of time delays $u_n$ for each element of the probe, an approximation equivalent to the phase screen model in the literature \cite{o1988phase, flax1988phase}. Although this model can account for a significant part of the aberration near the transducer, it is considered too simple to properly model ultrasound propagation in the medium, particularly for transcranial imaging when the skull thickness is not negligible.

In its more general form, $\mathbf{u}$ is composed of delays $u_n(x,z)$ that need to be applied for each position $(x,z)$ and for each received channel. This would represent a solution space of $N_x \times N_z \times N_{ele}$ parameters to optimize, which could limit convergence and increase the risks of overfitting. However, aberration generally remains invariant over small regions called isoplanatic patches that are larger than a single pixel. The size of a single isoplanatic patch varies, but usually covers multiple wavelengths \cite{bendjador2020svd, lambert2022bultrasound, robin2023vivo}. Hence, $\mathbf{u}$ can be parametrized with $\mathbf{u_p}$, the aberration function corresponding to the coordinates $(x_p,z_p)$ of the $N_p$ spatially distributed isoplanatic patches of the medium instead of each pixel location, which significantly reduced the size of the solution space needed to be optimized. More specifically, each $u_n(\mathbf{r})$ delay of $\mathbf{u}$ can be obtained by interpolating $\mathbf{u_p}$
\begin{align}
    \mathbf{u} = p(\mathbf{u_p}),
\end{align}

where $p(.)$ stands for an arbitrary smooth interpolation function acting as a regularizer. Gradient ascent can then be rewritten as
\begin{align}
    \mathbf{u_p}^{(j+1)} =     \mathbf{u_p}^{(j)} + \gamma\nabla_{\mathbf{u_p}} f(p(\mathbf{u_p}^{(j)})),
\end{align}

with a regularized solution space of $N_p\times N_{ele}$ parameters.

\subsection{Generalization to 3D beamforming}

Differentiable beamforming is a general approach to aberration correction and can also be extended to 3D imaging. Considering a matrix-array probe, plane waves can be emitted in both the $x$ and $y$ directions. The travel times can then be computed as
\begin{align}
    \tau_n(\mathbf{r}) &= \frac{x\sin\theta_x+y\sin\theta_y+z\sqrt{1-\sin^2\theta_x-\sin^2\theta_y}}{c}\\
    &+ \frac{\sqrt{z^2+(x-x_n)^2+(y-y_n)^2}}{c},
\end{align}
with $\theta_x$ and $\theta_y$ the plane wave angles in the $x$ and $y$ directions, respectively, and $(x_n,y_n)$ the element position of the matrix array. The aberration function is then defined as a set of 2D time delays that cover each element of the probe. The angular covariance matrix was computed for each depth. Furthermore, due to the higher sparsity and higher noise level of 3D ultrasound data, a 20 dB threshold was used to compute the covariance matrix.

To achieve a smooth 2D aberration profile, the second derivative regularization term was replaced with the 2D discrete laplacian operator. Spatial regularization was applied using the 27-point discrete 3D Laplacian operator
\begin{align}\mathbf{D}^2_{(x,y,z)}=
\frac{1}{26}
 \begin{bmatrix}
    \begin{bmatrix}
    2 &3 & 2\\
    3 & 6  & 3\\
    2 & 3 & 2
    \end{bmatrix},
    \begin{bmatrix}
    3 &6 & 3\\
    6 & -88  & 6\\
    3 & 6 & 3
    \end{bmatrix},
    \begin{bmatrix}
    2 &3 & 2\\
    3 & 6  & 3\\
    2 & 3 & 2
    \end{bmatrix}
        \end{bmatrix}.
\end{align}

\subsection{Ethics}

All experimental procedures were approved by the Animal Care Ethics Committee of the University of Montreal (permit numbers 21-017 and 22-013 for experiments in mice and permit numbers 20-085 and 21-065 for experiments in macaques), and followed the guidelines of the Canadian Council on Animal Care, and the Animal Research: Reporting of In Vivo Experiments (ARRIVE).

\subsection{ULM data acquisition}

\paragraph*{Acquisition in mice.} 

Seven 2-3 months old mice and four 6-7 month-old mice were used in this study. For all mice, excess hair was removed from the head, with the skin and skull kept intact. A 3D-printed head mount was used to limit movements from the mouse head during acquisition. Mice were kept under general anesthesia with ketamine (50 mg/kg) and medetomidine (1 mg/kg) and body temperature maintained between 36 and 37 °C with a heating pad. Before each acquisition, a 10 $\mu$L bolus of Definity microbubbles (Perflutren Lipid Microsphere, Lantheus Medical Imaging, Billerica, MA, USA) was injected in the retro-orbital sinus.

Ultrasound imaging was performed using a 128-element 15.625-MHz probe (L22-14, Vermon, France) connected to a Vantage 256 system (Verasonics Inc., Redmond, WA) as previously described \cite{xing2024phase, xing2025inverse}.  Ultrasound data were acquired continuously at a 100$\%$ sampling rate using 11 compounding angles (-5$^\circ$ to 5$^\circ$ with step of 1$^\circ$) with a 25 V transmit voltage. The pulse shape was composed of three cycles of a sinusoidal wave emitted at the central frequency of 15.625-MHz. A complete acquisition consisted of 600 buffers composed of 500 images acquired at a frame rate of 1000 Hz or 1200 buffers of 400 images acquired at a frame rate of 1600 Hz, for a total acquisition time of 5 minutes.

\paragraph*{Acquisition in macaques.}

Transcranial ultrasound imaging was performed in two adult Rhesus macaques under aseptic conditions and general anesthesia as previously described \cite{xing20253d}. Briefly, anesthesia was induced with an injection of ketamine (10 mg/Kg, intramuscular (IM)), given with glycopyrrolate (0.01 mg/Kg, IM) to prevent excessive salivation. Anesthesia was maintained with 2-3$\%$ isoflurane (Furane, Baxter) in 100$\%$ oxygen after tracheal intubation. Macaques were positioned in a stereotaxic frame to limit motion from the head. Vital signs (heart rate, respiratory rate, and blood oxygen saturation) were monitored during the entire experiment. A self-regulated heating pad (Harvard Apparatus, Holliston, MA, USA) was used to maintain temperature between 36.5-37 $^o$C. The ultrasound probe was manually positioned and held using a 3-axis clamp. Definity microbubbles (Lantheus Medical Imaging, Billerica, MA, USA) were injected through a peripheral venous catheter. Bolus injections of 100 µL microbubbles were used and flushed with a 0.5 mL phosphate buffered saline (PBS) solution.

Ultrasound imaging was performed using a 64-element 3-MHz phased array (P4-2, ATL/Philips, Andover, MA, USA) connected to a Vantage 256 system (Verasonics Inc., Redmond, WA). Ultrasound data were acquired continuously at a  100$\%$ sampling rate using 11 divergent waves from virtual sources (placed 10.24 mm behind the probe tilted by increment of 1$^\circ$) with a 30 V transmit voltage. The pulse shape was composed of three cycles of a sinusoidal wave emitted at the central frequency of 2.5-MHz. A complete acquisition consisted of 900 buffers composed of 500 images acquired at a frame rate of 1000 Hz,  for a total acquisition time of 7.5 minutes.

%For the P4-1 probe, ultrasound data were acquired continuously at a  100$\%$ sampling rate using 9 virtual sources (placed 24.53 mm behind the probe with tilted by increment of 1$^\circ$) with a 40 V transmit voltage. The pulse shape was composed of three cycles of a sinusoidal wave emitted at 1.5625-MHz. A complete acquisition consisted of 1200 buffers composed of 500 images acquired at a frame rate of 500 Hz, for a total acquisition time of \hl{20} minutes. 

\subsection{fUS imaging acquisition}

Three additional 8-9 week-old mice were used for the fUS imaging. The skin was removed and a 3D-printed head mount was implemented to limit head motion during experiments. fUS acquisitions of whisker stimulation were performed through the intact skull bone using a 128-element 15.625-MHz probe (L22-14, Vermon, France) connected to a Vantage NXT system (Verasonics Inc., Redmond, WA). Ultrasound data were continuously acquired using a 50 $\%$ sampling rate. The sequence consisted of 11 angles (-10$^\circ$ to 10$^\circ$ with step of 2$^\circ$) emitted at a pulse repetition frequency (PRF) of 5500 Hz with a transmit voltage of 25 V, for a compounded frame rate of 500 Hz. A stimulation cycle was composed of 10 s of pre-stimulation period (baseline), a 30 s of whisker stimulation followed by a 20 s of post-stimulation period (rest), for a total trial duration of 60 s. A complete fUS acquisition was composed of 5 trials for a total duration of 5 minutes.

\subsection{3D ULM data acquisition in nonhuman primates}

All procedures were conducted under aseptic conditions and general anesthesia, as previously described for 2D imaging. Experiments were performed in two healthy adult Rhesus macaques. A 3-MHz multiplexed $32\times32$ matrix-array probe of 1024 elements (Vermon, Tours, France) was used for 3D imaging. Acquisitions in macaque 1 were performed with the probe connected to a 256-channel Vantage system, and acquisitions in macaque 2 were performed with the probe connected to a Vantage NXT system. The sequence was composed of five plane waves in the x-axis $\{$(-2$^o$, 0$^o$),  (-1$^o$, 0$^o$),  (0$^o$, 0$^o$),  (-1$^o$, 0$^o$),  (2$^o$, 0$^o$)$\}$ emitted at 25 V with the full-aperture at the maximal pulse repetition frequency (PRF). Ultrasound data were acquired at a volume rate of 500 Hz in macaque 1 for a total of 600 blocks of 400 volumes and continuously acquired at a volume rate of 400 Hz in macaque 2 for a total of 300 blocks of 600 volumes.

\subsection{Data preprocessing and differentiable beamforming setting}

A bandpass singular value decomposition (SVD) clutter filter \cite{demene2015spatiotemporal} was applied to raw channel IQ data to remove tissue signals and clutter generated from the skull, as well as to improve noise suppression.  As ultrafast acquisitions are composed of multiple frames, differentiable beamforming was applied on temporal ensemble using the batch dimension of the PyTorch framework for the frames. To reduce computational load, a down-sampling factor between 25-50 was applied on the frame rate.

For data acquired in mice, an isometric beamforming grid of $\lambda/2 \times \lambda/2$ was used. The aberration function $\mathbf{u}$ was parametrized with isoplanatic patch sizes of $8\lambda \times 8\lambda$ with a field of view of $16 \times 10$. For data acquired in the Rhesus macaque brain using a phased array, a polar beamforming grid ($r, \theta$) was used with $dr=\sfrac{\lambda}{3}$ and $d\theta=0.5^\circ$. The field of view was separated into $3 \times 3$ patches of dimension of $30^\circ \times 26.7\lambda$. For 3D data, a beamforming grid of $\lambda\times\lambda\times\lambda$ was used. The field of view was then separated into $5\times5\times 12$ isoplanatic patches.

The aberration function was then upsampled with a cubic interpolation scheme using the patch centroid coordinates $(x_p, y_p, z_p)$ to cover the entire beamformer grid. The speed of sound used for beamforming was set at 1540 m/s and the learning rate was dynamically managed by the Adam optimizer. The parameters used are presented in Table \ref{tab:parameters}.

\begin{table}[h!]
    \centering
        \caption{Parameters used for differentiable beamforming}
    \begin{tabular}{m{6em}|m{5em}  m{5em} m{5em}}
    \hline
         Parameters & 2D mouse & 2D NHP & 3D NHP\\
         \hline
         $\gamma$ & 0.2 & 0.2 & 0.5\\
         Weight decay & 0 & 0 & $1\times10^{-5}$\\
         $\alpha_1$ & $1\times10^{-5}$  &$1\times10^{-4}$ & $2\times10^{-7}$ \\
         $\alpha_2$ &$5\times10^{-4}$ & $1\times10^{-5}$ & $2\times10^{-8}$\\
         $\alpha_3$ & $5\times10^{-4}$ & $1\times10^{-5}$ & $2\times10^{-8}$\\
         \hline
    \end{tabular}
    \label{tab:parameters}
\end{table}
%The initial state $\mathbf{u}^{(0)}$ was set to 0 ns for every patch and the maximum number of iterations for gradient ascent was fixed to 30. The learning rate was dynamically managed by the Adam optimizer.% and initially set heuristically at 0.2. %The regularization parameter $\alpha$ was set to $10^{-5}$ for 2D and $5\times10^{-8}$ for the 3D.

\subsection{ULM processing}

SVD filtering was performed on temporal ensembles composed of 400 or 500 frames. After beamforming, an adaptive spatially-dependent time gain compensation (TGC) map was computed from the Doppler signal using a Gaussian filter, as previously described \cite{xing20253d}. The TGC map was used to normalize the image intensity to improve signals in shadowed regions
\begin{align}
        \tilde{s}_(x,y,z,t) = \dfrac{s(x,y,z,t)}{TGC(x,y,z)}.
\end{align}

A lag-1 auto-correlation and a coherent temporal ensemble averaging with a Hanning window $w(t)$ of 5 frames were performed on IQ beamformed data to further enhance microbubble signals \cite{xing20253d}
\begin{align}
S(x,y,z,t) =\sqrt{\Big|\sum_{t-2}^{t+2} w(t)\cdot\tilde{s}_(x,y,z,t)\cdot \tilde{s}^*(x,y,z,t+1)\Big|}.
\end{align}

A square-root operation was applied on the auto-correlation to readjust signal dynamic range.

For mice, the tracking and localization algorithm was applied \cite{leconte2024tracking} using a 3D space-time Hessian-based vesselness filter to detect microbubble tracks. For macaque acquisitions with a phased-array, microbubbles were first localized on the polar grid using a weighted least-squares Gaussian fitting \cite{guo2011simple} and positions were then converted into Cartesian coordinates. 3D ULM in the macaque was processed as previously described \cite{xing20253d}. Only microbubbles with a local signal-to-noise ratio (SNR) higher than 20 dB were considered as true detections. Selected microbubbles were then tracked using an algorithm based on the Hungarian method \cite{heiles2022performance}. Each individual track was smoothed with an algorithm based on least-squares methods \cite{garcia2010robust} and then interpolated using the modified Akima method to retrieve continuous trajectories.

%To generate the vascular density maps, each track was projected and accumulated on a grid with a pixel size of $ \frac{1}{10}\times \frac{1}{10} \lambda^2$. The spatial interpolation step was matched to the projection grid to avoid gaps in vessel reconstruction, and each track was only counted once by pixel \cite{heiles2022performance}. 

%Only tracks longer than 15 frames were used for the ULM reconstruction. To improve ULM rendering, a 2D Gaussian filtering with standard deviation of 1 pixel was applied.

Velocities were computed with a forward finite difference scheme directly after smoothing and then interpolated using the modified Akima method \cite{xing20253d}. 

%After projection onto the reconstruction grid, velocities were normalized with the number of detected microbubbles \cite{heiles2022performance}. 
%To improve velocity maps final rendering, a binary mask was applied to remove velocity values in pixels where only a single microbubble was detected. A 2D Gaussian filtering with standard deviation of 1 pixel was also applied to further reduce noise level.

\subsection{fUS processing}

After beamforming, a bandpass SVD filtering was applied to remove tissue and noise signals. An additional Butterworth high-pass filter with a cut-off frequency of 50 Hz was also applied to further remove tissue or motion signals \cite{rungta2017light}. Each power Doppler image (PDI) was generated with the incoherent sum of 250 B-mode images. The PDIs were spatio-temporally smoothed using an Hanning window filter. Variations in relative cerebral blood volume ($\Delta$CBV) were computed by using the 10 s baseline as reference value. Correlation maps were computed using the Pearson's correlation coefficient $r$ between the stimulation pattern $A(t)$ and the $PDI(t)$ \cite{mace2011functional}
\begin{align}
    r (x,z)= \frac{\sum_{i=1}^{n_t}(PDI(t_i)-\overline{PDI})(A(t_i)-\overline{A})}{ \sqrt{\sum_{i=1}^{n_t}(PDI(t_i)-\overline{PDI})^2}\sqrt{\sum_{i=1}^{n_t}(A(t_i)-\overline{A})^2}},
\end{align}
where $\overline{PDI}$ and $\overline{A}$ denote the temporal mean values. Activation maps and activation time course were generated using the average signals over the 5 trials.

\subsection{Evaluation metrics}

To evaluate improvement in focusing, we also computed the spatial coherence using the coherence factor (CF) \cite{mallart1994adaptive}. The CF represents the ratio between the coherent and the incoherent sum of the time-delayed receive signal $s(.)$, such as 
\begin{align}
    CF_{(x,y)}(\mathbf{u}) = \frac{1}{N_{ele}N_{tx}}\frac{\big|\sum_{i=1}^{N_{tx}}\sum_{n=1}^{N_{ele}} s_{n,i}(x,z;\mathbf{u}) \big|^2}{\sum_{i=1}^{N_{tx}}\sum_{n=1}^{N_{ele}} |s_{n,i}(x,z; \mathbf{u})|^2} .
\end{align}

The CF is defined between 0 and 1. The optimal value of the CF for a perfectly coherent signal is 1 and its significantly decreased for incoherent sources such as speckle. The main advantage of the CF is that it is independent of the pixel intensity and depends only on the spatial coherence of the ultrasound signal along the aperture of the transducers.

To evaluate power Doppler images improvement after aberration correction, we measured the  contrast-to-noise ratio (CNR) and signal-to-noise ration (SNR)
\begin{align}
    CNR &= 10 \log_{10}\left(\frac{|\mu_{s}-\mu_{n}|}{\sigma_{n}}\right)\\
    SNR &= 10\log_{10} \left(\frac{\mu_{s}}{\sigma_{n}}\right),
\end{align}

where $\mu_{s}$ is the mean intensity within a region of interest, $\mu_{n}$, and $\sigma_{n}$ the mean intensity and standard deviation of noise, respectively.

\iffalse
To evaluate the similitude of aberration functions recovered at different concentrations, we previously introduced a modified cosine similarity metric \cite{xing2025inverse}. The similarity between aberration functions $\mathbf{u}_1$ and $\mathbf{u}_2$ can be computed as
\begin{align}
    C = \frac{\mathbf{u}_1 \cdot \mathbf{u}_2}{\max{\left(||\mathbf{u}_1||^2, ||\mathbf{u}_2||^2\right)}},
\end{align}

where $\cdot$ represents the dot product. The maximal norm of $\mathbf{u}_1$ or $\mathbf{u}_2$ was used as normalization factor to account for difference in phase amplitudes or offsets between the aberration functions. 
\fi

The Fourier ring correlation (FRC) \cite{hingot2021measuring} and the Fourier shell correlation (FSC) were used to evaluate the resolution of ULM in 2D and in 3D, respectively. Tracks were first randomly separated into two sub-images and the FRC/FSC was computed by using correlation between the spatial spectrum of each sub-image for pixels within a radius $r$
\begin{align}
    FRC(r) = \frac{\sum_{r\in R}F_1(r)\cdot F_2(r)^*}{\sqrt{\sum_{r\in R}|F_1(r)|^2\cdot \sum_{r\in R}|F_2(r)|^2}}.
\end{align}

The resolution was defined as the intersection of the FRC/FSC curve with the half-bit threshold. It is important to note that the FRC/FSC k-space extent reflects the effective spatial frequency support of the reconstructed vasculature, which increases when multiple microbubbles densely and diversely sample a structure, but this should also be interpreted as a sampling-dependent effective resolution rather than an absolute one.

\subsection{Statistical analysis}

All statistical analyzes were performed with MATLAB. For comparison across two groups that are pairs (meaning when the measurement is performed twice in the same subject), a two-tailed paired Student's \textit{t}-test was used, to evaluate statistical significance before and after correction for metrics such as contrast and resolution. For comparison across more than two groups, a repeated-measures one-way ANOVA was used when data points were paired across one factor, such for comparison of CF and $\mathbf{C}_\theta$ magnitude for different patch sizes or at different iteration time points of gradient ascent.

A repeated-measures two-way ANOVA was used when two factors were compared with data that were paired across one of the factors, such for comparing gain in contrast or change in vessel width before and after correction for different microbubble concentrations. A two-way ANOVA with post-hoc Student's \textit{t}-test and Bonferroni correction was used to evaluate the effect of two independent, categorical variables, such for establishing statistical significance by depth of the aberration function. The two-sample Kolmogorov-Smirnov test was used to evaluate statistical significance between distributions of large sample size data, such as for track velocity distributions and track length distributions. When applicable, data are displayed as the mean $\pm$ standard deviation. 

For parametric tests such as the Student's \textit{t}-test, data were assumed to be normally distributed. Unless specified otherwise, a statistical significance level of 5$\%$ was used. Levels of significance are given as follow :* $P < 0.05$, **  $P< 0.01$, ***  $P< 0.001$, and ****  $P< 0.0001$. When Bonferroni correction was used, the levels of significance were adjusted accordingly to the number of multiple comparisons performed.

\subsection{Computational resources}

Differentiable beamforming processing was performed on a working station with an Intel Xeon W-2245 3.9 GHz processor, 256 Go of RAM, and an NVIDIA GeForce RTX 3090 GPU with 24 Go of dedicated memory. The computational complexity of differentiable beamforming depends on some key parameters such as the beamforming grid sampling size and the number of iterations. With the configuration used in this study ($\lambda/2$ grid size and 30 iterations), an aberration function can be retrieved in approximately 2.2 minutes of computation. However, further optimization of the implementation, particularly optimizing PyTorch for GPU computing, could help reduce the computational time.

ULM processing was performed on high-performance compute (HPC) servers. Beamforming and localization were performed on a GPU accelerated system using an Intel Gold 6148 Skylake 2.4 GHz processor, 64 Go of RAM, and an Nvidia V100SXM2 GPU with 16 Go of dedicated memory. Tracking was performed on CPU only system using 10 cores (Intel Gold 6148 Skylake 2.4 GHz processor) and 128 Go of RAM.

\section{Data availability}

Part of the data acquired on mice are made publicly available as part of the ULMshare dataset (\url{https://www.frdr-dfdr.ca/repo/collection/tulmshare}). Additional data from this study are available from the corresponding author upon reasonable request.

\section{Code availability}

An example code will be made available upon publication.

\section{Acknowledgment}

This work was supported by the New Frontier in Research Fund under grant NFRFE-2022-00590, the Canada Foundation for Innovation under Grant 38095, the Canadian Institutes of Health Research (CIHR) under Grant 452530, PJT-156047 and MPI-452530, the Natural Sciences and Engineering Research Council of Canada (NSERC) under discovery grants RGPIN-2020-06786 and RGPIN-2020-05276, by the Canada Research Chair in neurovascular interactions, and by Brain Canada under grant PSG2019. Computer supports were provided by the Digital Research Alliance of Canada. P.X. was was supported by the TransMedTech Institute and by NSERC. A.M. was supported by the Mitacs accelerate fellowship.

\section{Author contributions}
P.X. developed the aberration correction method and the ultrasound acquisition protocol. P.X., A.M., E.M., and S.Q. performed experiments. P.X. analyzed the data, conceived the figures, and wrote the first version of the manuscript. J.P., R.R., N.D. supervised the study. 

\section{Competing interests}
J.P. is a member of the Verasonics Scientific Advisory Board.

\bibliography{references}

\newpage

%\section{Supplemental materials}\label{section:appendix}

\renewcommand{\thefigure}{S\arabic{figure}}
\setcounter{figure}{0}

\begin{figure*}
    \centering
    \includegraphics[width=1\linewidth]{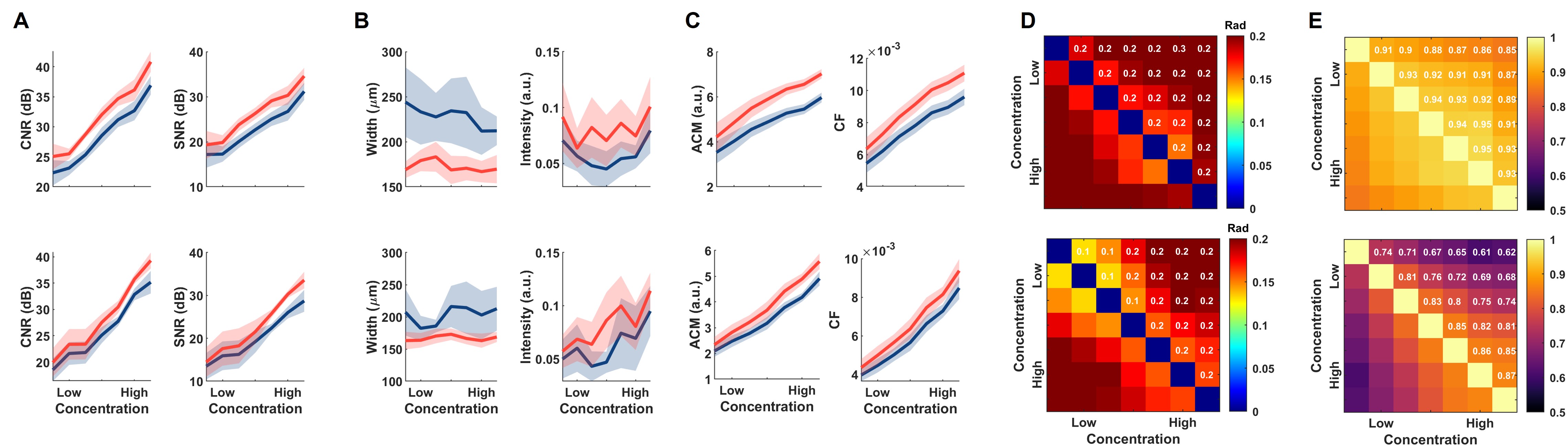}
    \caption[Microbubble concentration effect on correction performance]{Microbubble concentration effect on correction performance. (\textbf{A}) Increase in CNR and SNR as a function of microbubble concentrations in young and old mice. CNR increased with microbubble concentration (young mice: $n=7$, $F=13.2$, $P  < $ 0.0001, old mice: $n=4$, $F=9.2$, $P  < $ 0.001, repeated-measures two-way ANOVA) and improved after correction (young mice: $n=7$, $F=26.0$, $P  < $ 0.01, old mice: $n=4$, $F=12.0$, $P  < $ 0.05, repeated-measures two-way ANOVA). Similarly, SNR increased with microbubble concentration (young mice: $n=7$, $F=9.4$, $P  < $ 0.0001, old mice: $n=4$, $F=7.2$, $P  < $ 0.001, repeated-measures two-way ANOVA) and improved after correction (young mice: $n=7$, $F=24.5$, $P  < $ 0.01, old mice: $n=4$, $F=10.2$, $P  < $ 0.05, repeated-measures two-way ANOVA).
    (\textbf{B}) Vessel width and vessel intensity after correction as a function of microbubble concentrations. Vessel width improved in young mice (young mice: $n=7$, $F=10.6$, $P  < $ 0.05, old mice: $n=4$, $F=1.8$, $P  = $ 0.27, repeated-measures two-way ANOVA) without effect of microbubble concentration (young mice: $n=7$, $F=0.2$, $P  = $ 0.97, old mice: $n=4$, $F=0.8$, $P  = $ 0.61, repeated-measures two-way ANOVA). Vessel intensity tended to increase after correction (young mice: $n=7$, $F=5.9$, $P  = $ 0.051, old mice: $n=4$, $F=1.6$, $P  = $ 0.30, repeated-measures two-way ANOVA) without effect of microbubble concentration (young mice: $n=7$, $F=0.8$, $P  = $ 0.57, old mice: $n=4$, $F=0.8$, $P  = $ 0.60, repeated-measures two-way ANOVA).
    (\textbf{C}) Increase in CF and $\mathbf{C}_\theta$ as a function of microbubble concentration. CF increased with microbubble concentration (young mice: $n=7$, $F=23.7$, $P  < $ 0.0001, old mice: $n=4$, $F=26.3$, $P  < $ 0.0001, repeated-measures two-way ANOVA) and improved after correction (young mice: $n=7$, $F=282.9$, $P  < $ 0.0001, old mice: $n=4$, $F=33.7$, $P  < $ 0.05, repeated-measures two-way ANOVA).  $\mathbf{C}_\theta$ increased with microbubble concentration (young mice: $n=7$, $F=21.3$, $P  < $ 0.0001, old mice: $n=4$, $F=30.4$, $P  < $ 0.05, repeated-measures two-way ANOVA) and improved after correction (young mice: $n=7$, $F=101.4$, $P  < $ 0.0001, old mice: $n=4$, $F=29.1$, $P  < $ 0.05, repeated-measures two-way ANOVA). 
    (\textbf{D}) Mean variation in phase for the aberration functions obtained at different microbubble concentrations. (\textbf{E}) Aberration function correlation coefficient for different microbubble concentrations.}
    \label{fig:concentration}
\end{figure*}

\begin{figure*}
    \centering
\includegraphics[width=0.6\textwidth]{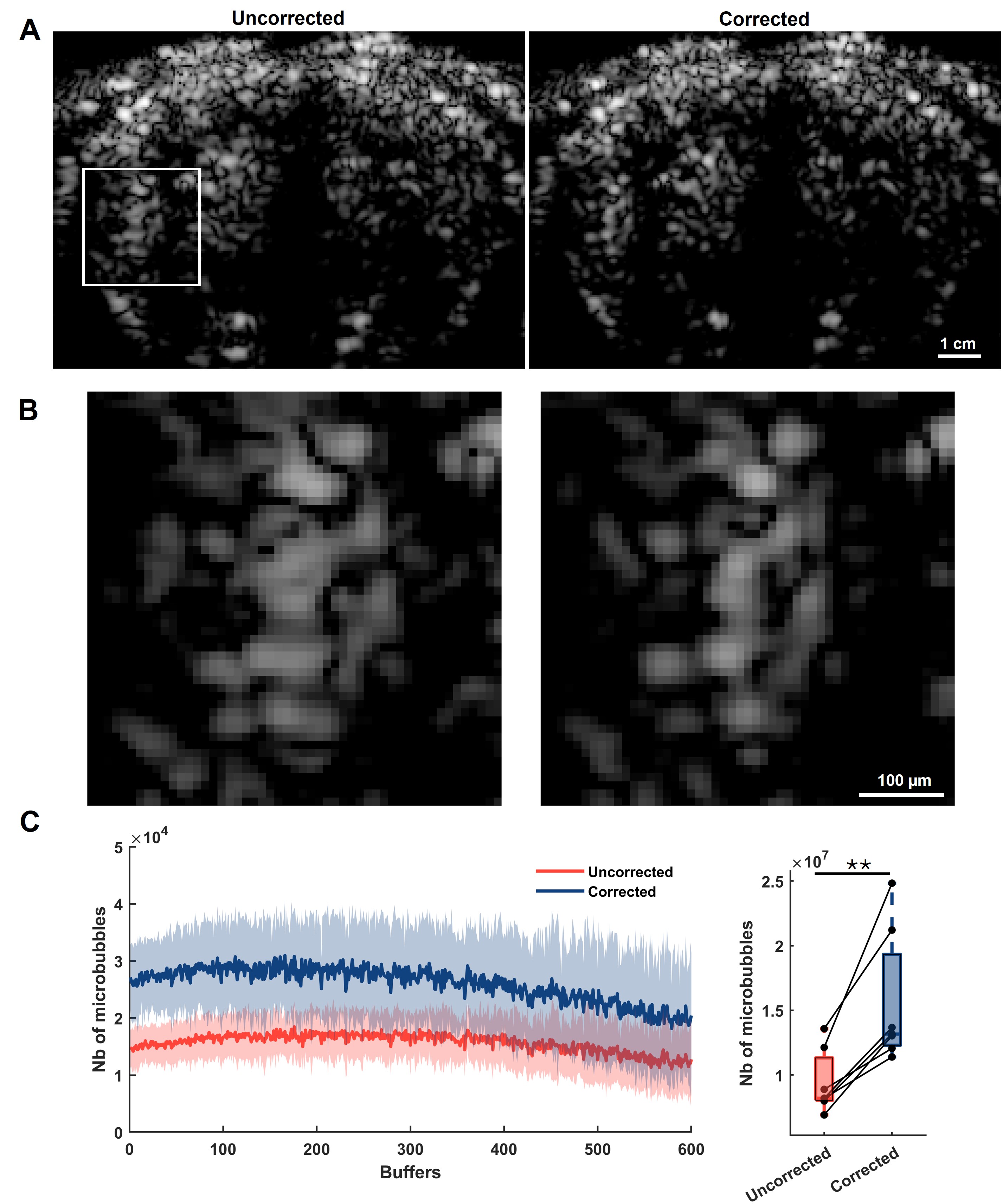}
    \caption[Differentiable beamforming enhances microbubble detection]{Differentiable beamforming enhances microbubble detection. (\textbf{A}) Example of Bmode before and after correction. (\textbf{B}) ROI showing microbubbles before and after correction. (\textbf{C}) Microbubble detection rate before and after correction. Microbubble detection increased from 9.41 $\pm$ $2.5\times10^6$ to  15.62 $\pm$ $5.2\times10^6$ after correction ($n=7$, $P  < $ 0.01, two-tailed paired Student's \textit{t}-test).}
    \label{fig:supp_microbubbles}
\end{figure*}

\begin{figure*}
    \centering
    \includegraphics[width=1\linewidth]{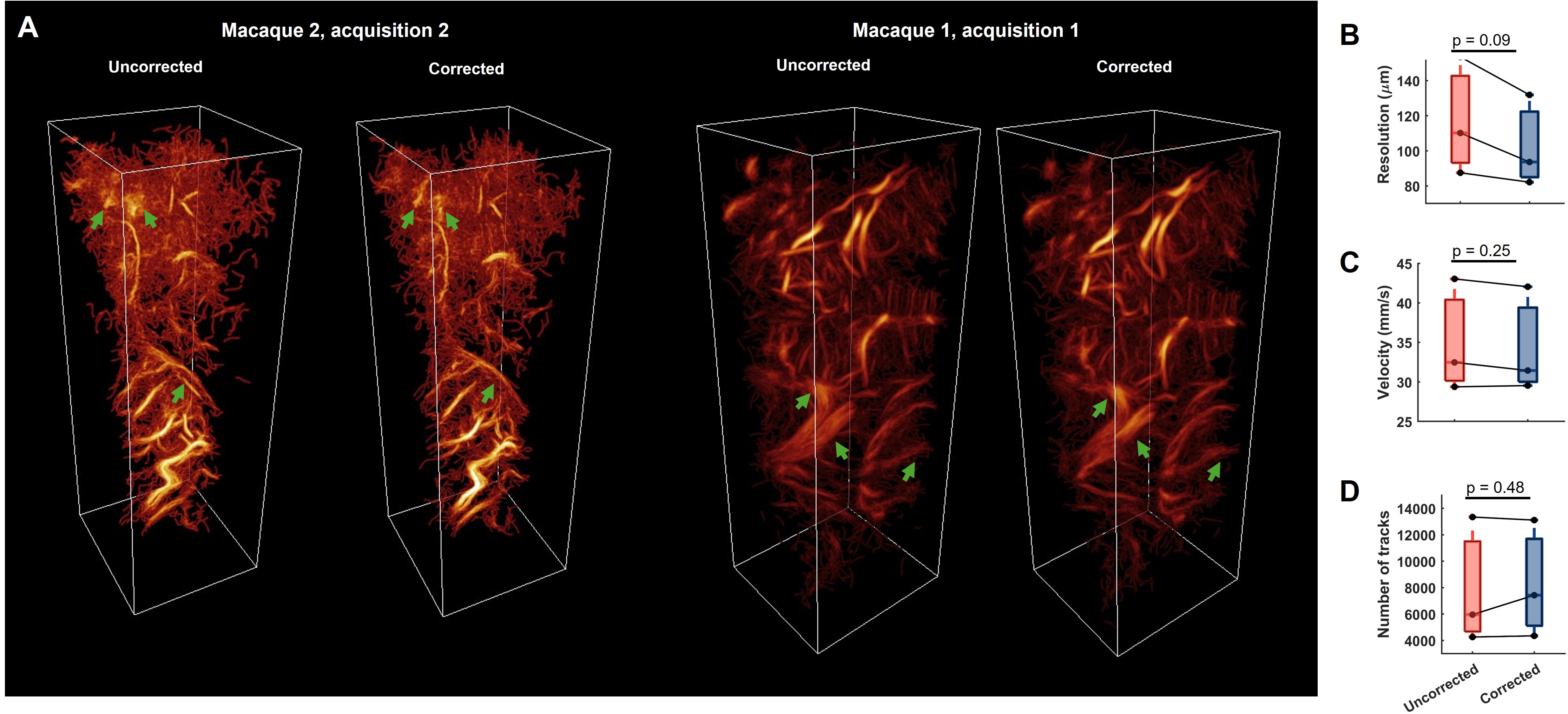}
    \caption{Additional transcranial 3D ULM in NHPs with differentiable beamforming. (\textbf{A}) ULM maps before and after correction from macaques 1 and 2. (\textbf{B}) FSC before and after aberration correction ($n=3$, $P=0.09$, paired Student's \textit{t}-test). (\textbf{C}) Mean track velocity before and after aberration correction ($n=3$, $P=0.25$, paired Student's \textit{t}-test). (\textbf{D}) Mean track length before and after aberration correction ($n=3$, $P=0.48$, paired Student's \textit{t}-test).}
    \label{fig:supp_3d}
\end{figure*}

\end{document}